\documentstyle[epsf,aps]{revtex}

\begin{document}
\title{Self consistent estimates of magnetic fields from reheating}
\author{Esteban Calzetta\footnote{e-mail: calzetta@df.uba.ar} and 
Alejandra Kandus\footnote{e-mail: kandus@iagusp.usp.br}\footnote{present
address: IAG - Universidade de Sao Paulo, S.P., Brasil}}
\address{Depto. de F\'{\i}sica, F. C. E. y N. - UBA
\\
 pab. I, Cdad. Universitaria
\\ 
(1428) Buenos Aires, Argentina}
\maketitle

\begin{abstract}
We investigate the generation of primordial magnetic fields from stochastic
currents created by the cosmological transition from inflation to reheating.
We consider N charged scalar fields coupled to the electromagnetic field in
a curved background and derive self-consistent equations for the evolution of
the two point functions of the fields, which in the large-N limit give a
decoupled set for the scalar and the electromagnetic functions.
The main contribution to the electric current comes from the infrared
portion of the spectrum of created particles, and in this limit 
the damping of the magnetic field is not due to normal conductivity but to 
London currents in the scalar field. 
For a given set of the physical parameters of the problem, 
we solved this equation numerically and found that, due to the fact that
the London currents are oscillating, the field actually grows 
exponentially during the time interval in which our large-N limit equations 
are valid. Although for the chosen parameters the induced field is weak, the 
present uncertainties on their actual values leave open the possibility for 
higher intensities.

\end{abstract}

\section{Introduction}

One of the most intriguing observational aspects of astrophysics is the
detection of large scale magnetic fields in all the cosmological structures
of the Universe \cite{oren,zweibel,ensslin,beck,zeldovich}. The remarkable
aspect of the observations of galactic magnetic fields, is that their 
intensity and structure are more or less the same, no matter if the galaxy
is one of our neighbors or a highly red-shifted one. Clusters of galaxies are
also permeated with intense and quite coherent magnetic fields.

The main puzzle is the dynamical origin of these large scale fields. For
galactic fields, two main lines of research are pursued: one proposes an
origin based on local astrophysical processes, while the other advocates a
primordial origin. The seducing aspect of a primordial origin of these
fields is that it accounts for the observations in all the mentioned
structures, no matter how high their red-shift is.

The main difficulty for generating magnetic fields in the Early Universe is
the need to break the conformal invariance of the Maxwell field. This can be
done in many ways, which explains the large variety of proposed mechanisms
in the literature \cite{harrison,quashnock,cheng,sigl,baym,widrow,spedalieri,dolgov,ratra,ale1,giovannini,lemoine,lemoine2,gasperini,maroto,opher}. After 
this first obstacle is overcome, the success of any given proposal
hinges on whether the energy from the primordial source may be successfully
stored in the magnetic field, or else it is dissipated away (over and above
the strong reduction brought by cosmic redshift). This point was highlighted
in a recent paper by Giovannini and Shaposhnikov \cite{misha1}. Using
kinetic theory, these authors showed that the amount of the magnetic field
which may be generated during reheating is highly sensitive to the
conductivity of the cosmic medium.

In this paper, we extend this analysis to fields generated in the earliest
stages of reheating. At this time the fields and their sources span
super-horizon scales and so they may hardly be described in terms of\ more or
less localized excitations, counted by a distribution function obeying a
transport equation. It becomes necessary to recast the analysis of \cite
{misha1} in terms of a more fundamental description of the process. We shall
atempt to obtain an upper bound for the magnetic field which may be
generated during reheating, directly from quantum field theory (for a
discussion of the circumstances in which quantum field theory reduces to
kinetic theory see \cite{esteban2}).

As a representative mechanism for field generation, we shall adopt the one
proposed earlier in ref. \cite{ale1,ale2}. In this proposal, conformal
symmetry is broken by coupling the Maxwell field to charged scalars fields.
The transition from inflation to reheating produces a strong amplification
of the scalar fields (particle creation). This results in stochastic
currents which eventually decay into the magnetic field. The main advantages
of this mechanism over others proposed in the literature is that it is
naturally superhorizoned and relatively simple, involving no new physics. 
The bulk of the field is
generated early in the reheating era (the so-called preheating period) when
on the other hand a quantum field theoretic treatment may be atempted.
It is realistic in the sense that there are actual candidates for the scalar
fields in supersymmetric versions of the standard model. We shall ignore the
difficulties associated with the fact that the electromagnetic field does
not exist as such during this period (we should rather follow the evolution
of the several vector fields in some unified theory, a combination of which
becomes the photon after electroweak symmetry breaking) as we believe our
order of magnitude estimates shall be validated by this more complete
treatment.

Given the present state of knowledge, a quantum field treatment can only be
done within some perturbative scheme. We shall assume the model contains $N$
identical scalar fields, and thus avails ourselves of the $1/N$ expansion 
\cite{jackiw,cooper}. Working at leading order, we shall obtain the
self-consistent evolution equations for mean fields and propagators using
the closed time path, two particle irreducible effective action \cite
{jackiw,esteban1}. These tools have been applied to study a broad spectrum
of non equilibrium problems, from heavy-ion collisions and pair production
in strong electric fields \cite{cooper} to the study of non equilibrium
dynamics of the inflaton field during reheating \cite{steve} and to the
evolution of quantum fields and production of density perturbation in
inflationary dynamics \cite{boyanovsky}

We shall study the evolution of a system of $N$ charged, massive scalar
fields coupled to one electromagnetic field, in an expanding universe which
realizes a transition from an exponential expansion (inflationary period) to
a less rapid expanding epoch (reheating period). For minimal coupling of the
scalar field to gravity, we can assume that during inflation the charged
field is in its invariant vacuum state \cite{allen}, there being no particle
creation and no electric currents.

When the transition from inflation to reheating takes place, the vacuum
state changes and the state of the field turns into a multi-particle state.
From then on, we have an out of equilibrium system of charged and
electromagnetic fields, that evolves in an expanding background, and it is
this evolution that we shall study.

The reheating epoch of the universe is very difficult to address, both from
the conceptual and from the technical point of view. Very little is
understood about the process of decay of the inflaton into the other fields
to which it is coupled, and the consequent establishment of the primordial
plasma \cite{brandenberger}. A complete description of the evolution of our
system of scalar particles and electromagnetic fields should take into
account their possible couplings to the other fields present. To atempt to 
take into account all these effects in a consistent
way, makes the problem conceptually and technically insolvable. However we
can try and incorporate them in a phenomenological way by
considering a ''thermal mass'' for the scalar particles, besides their bare
one. On the other hand, any other coupling to the forming plasma would
amount to consider an external electric conductivity, and this case 
was already considered in the literature \cite{ale1,misha1,ale2}.

We must observe that to preserve the very weak coupling of the inflaton
during inflation \cite{kolb}, it may be presumed that the inflaton is
neutral and uncoupled to other charged species. Therefore the plasma of
charged particles would not result from the decay of the inflaton, but from
some indirect mechanism (e. g., from gravitational particle creation \cite
{vilenkin}). As a result, the primordial charged plasma contributing to the
thermal mass of our scalar fields, need not be in equilibrium with the
neutral particles, including the inflaton, which contribute most to the
energy density of the Universe during reheating, and it will take longer to
form. As we shall assume that the primordial plasma is not established
immediately after the end of inflation, we shall consider no external
electrical conductivity in the period of time during which we follow
the evolution of our system of charges and magnetic field: the damping of
the fields and currents would be due to their own interaction.

The paper is organized as follows: In the next section we present the
general tools that we shall use in the magnetic field evaluation: we
evaluate in conformal time the scale factors of the Universe during
inflation and reheating; we write down and solve the Klein Gordon equation
for the scalar field modes in each epoch, and calculate the Bogolyubov
coefficients. As it is not possible find a closed analytical solution to 
the scalar field
equation during the reheating epoch, we split the momentum interval in two
parts: an infrared and an ultraviolet one and find the solutions for each of
them. The resulting Bogolyubov coefficients show that the bulk of the
particles are created in the infrared portion. In section III we write down
the Schwinger Dyson equation for the Hadamard two point function of the
electromagnetic field four potential, $D_{1\mu \nu }\left( x,x^{\prime
}\right) =\left\langle \left\{ A_{\mu }\left( x\right) ,A_{\nu }\left(
x\right) \right\} \right\rangle $ which will give us the information we seek
about the induced magnetic field. This is in fact a set of coupled equations
for the different components of the two point function. They posses two
kernels: a local one, and a non local one that can be associated with
dissipation. The main contribution to these kernels comes from the infrared
sector. In this situation, the equations for the transverse part of the pure
spatial components of the Hadamard function, i.e. the ones whose curl gives
the magnetic field, decouple. When computing the kernels, the dissipative
one turns out to be several orders of magnitude smaller than the local one
and hence can be ignored. The remaining local kernel will be responsible for
screening the field like in super-conducting media, i.e. Meissner effect.

In section IV we translate our equations to a stochastic formulation, which
allows a better understanding of the field evolution. We write down a
Langevin-like equation for the magnetic field itself, and
find a numerical solution for a given set of physical parameters of the
problem. Due to the fact that the local kernel is an oscillating
function of time we find that,
for all astrophysically interesting scales (i.e. the ones that
are superhorizoned during reheating), the field intensity grows
non adiabatically with the time, a behavior that can be understood in
terms of stochastic resonance \cite{fujisaki,kofman}.
Finally in section V we shortly review and discuss our results. 

In Appendix 1 we write down the main ingredients for our study, the
Lagrangian for scalar electrodynamics in curved space time, we build the 2PI
closed time path effective action and from it derive the set of
Schwinger-Dyson equations for the scalar and electromagnetic two point
functions. In taking the large {\em N} limit we find that the equation for
the scalar field propagator decouples from the one for the electromagnetic
field, and hence there is no back-reaction on the electric current. All the
information about the dissipative properties of the system is encoded in the
kernels of the equations for the electromagnetic two point functions. These 
equations are a set of coupled equations for the different components of the 
electromagnetic two point function.

In Appendix 2 we compute the kernels that appear in the Schwinger - Dyson
equations for the ultraviolet sector of the mode spectrum and for the
infrared one, and conclude that the latter dominates over the former. We
work in natural units, in which $c=\hbar =1$, and with signature $\left(
-,+,+,+\right) $.

\section{Scenario for Magnetogenesis}

Assume that there exists a charged massive scalar field in the very early
universe. Consider also de Sitter inflation, and that during that period of
the universe the scalar field is in its invariant vacuum state and hence
there is no particle creation \cite{allen}. When the transition from
inflation to reheating takes place, due to the change in the geometry of the
universe the vacuum state becomes a multiparticle state \cite{birrel}, which
means that a stochastic electric current unfolds which induces a magnetic
field.

As stated in the introduction, the reheating period is very difficult to
address. It is widely accepted that the general picture is the decaying of
the inflaton field into other matter fields through non linear oscillations.
This is an out of equilibrium process, during which the plasma that will
determine the evolution of the universe at the subsequent epochs is being
formed. It is usually assumed that it ends when (conformal) thermal
equilibrium is achieved, which in turn determines the beginning of the
radiation dominated epoch. Very little is understood about the formation of
the plasma, when the charged matter arises and how the temperature evolves 
\cite{brandenberger}

Our created particles will dive through a sea of forming matter fields and
it would be very naive to neglect a possible interaction between these two
systems, no matter how little we know about them. We shall consider that
this possible interaction produces a shift in the value of the scalar mass,
that can be much larger than its bare value. Due to the uncertainties
about the very process of particle creation (e.g. the change in the geometry
is not instantaneous and hence neither is particle creation) and about the
onset of reheating, we shall take into account this shift when calculating
particle creation. When studying the propagation of the induced magnetic
field, we shall not consider a possible conductivity of the forming plasma.
On one side this is a reasonable assumption for the early stages of
reheating, according to the comments made above. On other side this
assumption will allow us investigate the damping in the magnetic field due
to its own sources.

The scenario we are considering is then as follows: we evaluate particle
creation by matching the modes of the scalar field and their first time
derivatives at the instant of transition between inflation and reheating. We
assume that this instant is $\tau =0$ and that during inflation the universe
expands exponentially. In the early stages of reheating, the dominant form
of matter is still the oscillating inflaton field, and thus the universe
expands like under matter dominance \cite{starobinskii,kolb}. There being no
compelling argument to the contrary, we consider minimal coupling of the
matter field to the geometry. To model the generation of magnetic fields
during this period, we shall follow the interaction of the Maxwell field
with $N$ identical charged scalar fields, minimally coupled to the Maxwell
field. To leading order in the $1/N$ expansion, these fields obey the free
Klein-Gordon equation (see Appendix 1), and may be decomposed into modes in
the usual way. We shall assume geometry is described by a spatially flat
Friedmann-Robertson-Walker model.

\subsection{Scale Factors of the Universe}

To give a definite form to our model, we must find the scale factors of the
Universe for the inflationary and the reheating periods in conformal time,
which is defined as $d\eta =dt/a\left( t\right) $, $t$ being the physical or
cosmological time. For the inflationary epoch, the scale factor of the
Universe reads $a_{I}\left( t\right) =a_{0}\exp \left( Ht\right) $, so the
conformal time is $\eta -\eta _{0}=-\exp \left( -Ht\right) /Ha_{0}$ and in
this variable the scale factor reads $a_{I}\left( \eta \right) =H^{-1}\left(
\eta _{0}-\eta \right) ^{-1}$.

Assuming that during reheating the universe expands as if it were matter
dominated \cite{kolb}, in cosmological time the scale factor is $a_{R}\left(
t\right) =a_{1}\left( t+t_{1}\right) ^{2/3}$. Therefore the conformal time
reads $\eta -\eta _{1}=3\left( t+t_{1}\right) ^{1/3}/a_{1}$ and the
corresponding scale factor $a_{R}\left( \eta \right) =a_{1}^{3}\left( \eta
-\eta _{1}\right) ^{2}/9$.

In order to avoid overproduction of created particles, when matching the
scale factors we must demand that at the time of transition, the scale
factor changes smoothly. The minimum conditions to be fulfilled are that the
functions and their first derivatives be continuous at the transition time,
then obtaining $a_{I}\left( \eta \right) =\left( 1-H\eta \right) ^{-1}$ and $%
a_{R}\left( \eta \right) =\left( 1+H\eta /2\right) ^{2}$. Defining the
dimensionless time $\tau \equiv H\eta $ the corresponding functions read $%
a_{I}\left( \tau \right) =\left( 1-\tau \right) ^{-1}$ and $a_{R}\left( \tau
\right) =\left( 1+\tau /2\right) ^{2}$.

\subsection{Field Equations and Bogolyubov coefficients.}

In this section we solve the Klein Gordon equation for the scalar field, in
the two eras of the Universe involved in our study and calculate the
Bogolyubov coefficients. From now on we will work with dimensionless
variables $\kappa =Hk,\quad m\rightarrow m/H$, together with the
dimensionless time variable $\tau $ defined above.

Assuming minimal coupling to curvature the Klein Gordon equation for the
modes of the scalar field reads 
\begin{equation}
\left[ \frac{d^{2}}{d\tau ^{2}}+k^{2}+\frac{m^{2}\left( \tau \right) }{H^{2}}%
a^{2}\left( \tau \right) -\frac{\ddot{a}\left( \tau \right) }{a\left( \tau
\right) }\right] f_{k}\left( \tau \right) =0  \label{waq}
\end{equation}
In spite of our ignorance about the reheating epoch and the actual
interactions of our scalar field during it, it would be very naive to
neglect them, so as stated in the Introduction, we shall take them into
account by considering a thermal mass given by the phenomenological
expression 
\begin{equation}
m^{2}\left( \tau \right) =gT^{2}\left( \tau \right)  \label{war}
\end{equation}
with $g$ being a coupling constant which we consider of order one. We assume
a generic form for the evolution of the thermal mass during reheating, i.e.
one of the form

\begin{equation}
T\left( \tau \right) =\frac{a^{b}\left( 0\right) }{a^{b}\left( \tau \right) }%
T_{M}  \label{haa}
\end{equation}
where $b$ is a parameter that satisfies $0\leq b<1$, and $T_{M}$ is the
maximum value of the shift, attained at $\tau \simeq 0$. This form of the
mass shift is inspired in the process of preheating, where a peak
temperature is attained which subsequently decreases as reheating unfolds.
Setting $b=0$, $g=1$ implies $T_{M}=m_{0}$, where $m_{0}$ is the bare mass
of the scalar field. For any other values of $b>0$ it must be $T_{M}>m_{0}$.

The Bogolyubov coefficients are given by the decomposition 
\begin{equation}
f_{k}\left( \tau \right) =\alpha _{k}f_{k}^{R}\left( \tau \right) +\beta
_{k}f_{k}^{R\ast }\left( \tau \right)  \label{haaa}
\end{equation}
with $f_{k}^{R}\left( \tau \right) $ the modes of the scalar field during
reheating, and can be obtained by demanding continuity of the mode functions
and their derivatives at the time of transition.

\subsubsection{Inflation}

During this period, as the temperature of the Universe is practically zero,
the modes satisfy the bare mass equation, i.e. 
\begin{equation}
\left[ \frac{d^{2}}{d\tau ^{2}}+k^{2}-\frac{2-m_{0}^{2}/H^{2}}{\left( 1-\tau
\right) ^{2}}\right] f_{k}\left( \tau \right) =0  \label{wau}
\end{equation}
Writing $f_{k}\left( \tau \right) =\left( 1-\tau \right) ^{1/2}h_{k}\left(
\tau \right) $ we obtain a Bessel equation for $h_{k}\left( \tau \right) $: 
\begin{equation}
\left[ \frac{d^{2}}{d\tau ^{2}}-\frac{1}{\left( 1-\tau \right) }\frac{d}{%
d\tau }+k^{2}-\frac{9/4-m_{0}^{2}/H^{2}}{\left( 1-\tau \right) ^{2}}\right]
h_{k}\left( \tau \right) =0  \label{waw}
\end{equation}
whose positive frequency solutions are the Hankel functions of the first
kind, $H_{\nu }^{\left( 1\right) }\left[ k\left( 1-\tau \right) \right] $
with $\nu =\left( 3/2\right) \sqrt{1-4m_{0}^{2}/9H^{2}}\sim 3/2$; the last
relation stems from the fact that during inflation $m_{0}^{2}/H^{2}\ll 1$.
We then have that the normalized, positive frequency modes during inflation
read 
\begin{equation}
f_{k}^{I}\left( \tau \right) \simeq \frac{\sqrt{\pi }}{2}\left( 1-\tau
\right) ^{1/2}H_{3/2}^{\left( 1\right) }\left[ k\left( 1-\tau \right) \right]
=-\frac{e^{ik\left( 1-\tau \right) }}{\sqrt{2k}}\left[ 1+\frac{i}{k\left(
1-\tau \right) }\right]  \label{waz}
\end{equation}
The last equality follows from the exact polynomial expression for $%
H_{3/2}^{\left( 1\right) }\left[ k\left( 1-\tau \right) \right] $ \cite
{abramowitz}.

\subsubsection{Reheating}

For this epoch the Klein Gordon equation reads 
\begin{equation}
\left[ \frac{d^2}{d\tau ^2}+k^2+g\frac{T_M^2}{H^2}\left( 1+\frac \tau 2%
\right) ^{4\left( 1-b\right) }-\frac 1{2\left( 1+\tau /2\right) ^2}\right]
f_k\left( \tau \right) =0  \label{waqq}
\end{equation}

In this case it has no closed analytic solution, so we shall solve it in two
limits, namely the large wavenumber limit, for which $k\gg \Delta \equiv
g^{1/2}T_{M}/H$ and the small wavenumber limit, for which $k\ll \Delta $.
Care must be taken when doing this, because in eq. (\ref{waqq}) the factor $%
g^{1/2}T_{M}/H$ is multiplied by a growing functions of time. This means
that in principle this splitting is time dependent. However this will prove
to be unimportant at the end of the calculations.

\paragraph{Large wavenumber limit}

The Klein Gordon equation in this limit reads 
\begin{equation}
\left[ \frac{\partial ^{2}}{\partial \tau ^{2}}+k^{2}-\frac{1}{2\left(
1+\tau /2\right) ^{2}}\right] f_{\left( l\right) k}^{R}\left( \tau \right) =0
\label{xaa}
\end{equation}
proposing again $f_{\left( l\right) k}^{R}\left( \tau \right) =\left( 1+\tau
/2\right) ^{1/2}h_{R\left( l\right) k}\left( \tau \right) $ we obtain a
Bessel equation for $h_{k}\left( \tau \right) $: 
\begin{equation}
\ddot{h}_{R\left( l\right) k}\left( \tau \right) +\frac{1}{\left( 1+\tau
/2\right) }\dot{h}_{R\left( l\right) k}\left( \tau \right) +\left[ 4k^{2}-%
\frac{9/4}{\left( 1+\tau /2\right) ^{2}}\right] h_{R\left( l\right) k}\left(
\tau \right) =0  \label{xac}
\end{equation}
whose positive frequency solutions are the Hankel functions of the
second kind $H_{3/2}^{\left( 2\right) }\left[ 2k\left( 1+\tau /2\right) %
\right] $. The normalized mode functions read 
\begin{eqnarray}
f_{\left( l\right) k}^{R}\left( \tau \right) &=&\sqrt{\frac{\pi }{2}}\left(
1+\frac{\tau }{2}\right) ^{1/2}H_{3/2}^{\left( 2\right) }\left[ 2k\left( 1+%
\frac{\tau }{2}\right) \right]  \nonumber \\
&=&-\frac{e^{-i2k\left( 1+\tau /2\right) }}{\sqrt{2k}}\left\{ 1-\frac{i}{%
2k\left( 1+\tau /2\right) }\right\}  \label{xad}
\end{eqnarray}
where in the last line we have used the polynomial expression of the Hankel
function.

Replacing the corresponding modes and their derivatives in the general
expression for the Bogolyubov coefficients we get 
\begin{equation}
\alpha _{\left( l\right) k}=e^{i3k}\left[ 1+\frac{3i}{2k}-\frac{9}{8k^{2}}-%
\frac{3i}{8k^{3}}\right]  \label{xae}
\end{equation}
\begin{equation}
\beta _{\left( l\right) k}=e^{-ik}\left[ \frac{3}{8k^{2}}-\frac{3i}{8k^{3}}%
\right]  \label{xaf}
\end{equation}

\paragraph{Small wavenumber limit}

For small wavenumbers, i.e. those for which $k\ll g^{1/2}T_{M}/H$, the
equation reads 
\begin{equation}
\left[ \frac{d^{2}}{d\tau ^{2}}+g\frac{T_{M}^{2}}{H^{2}}\left( 1+\frac{\tau 
}{2}\right) ^{4-4b}-\frac{1/2}{\left( 1+\tau /2\right) ^{2}}\right]
f_{\left( s\right) k}\left( \tau \right) =0  \label{wbb}
\end{equation}

Assuming as time variable $u=\left( 1+\tau /2\right) $ and replacing $%
f_{\left( s\right) k}^{R}\left( z\right) =u^{1/2}h\left( u\right) $ we get 
\begin{equation}
\ddot{h}_{\left( s\right) k}\left( u\right) +\frac{1}{z}\dot{h}_{\left(
s\right) k}\left( u\right) +\left[ 4g\frac{T_{M}^{2}}{H^{2}}u^{c}-\frac{9/4}{
u^{2}}\right] h_{\left( s\right) k}\left( u\right) =0  \label{wbg}
\end{equation}
with $c=4-4b.$ Defining $x=u^{\gamma }$ the equation now reads 
\begin{equation}
\frac{d^{2}}{dx^{2}}h_{\left( s\right) k}\left( x\right) +\frac{1}{x}\frac{d%
}{dx}h_{\left( s\right) k}\left( x\right) +\left[ 4g\frac{T_{M}^{2}}{H^{2}}%
\frac{x^{c/p-2+2/p}}{\gamma ^{2}}-\frac{9}{4\gamma ^{2}x^{2}}\right]
h_{\left( s\right) k}\left( x\right) =0  \label{wbi}
\end{equation}
which can be cast in the form of a Bessel equation by demanding $c/\gamma
+2/\gamma -2=0$. We get $\gamma =\left( 3-2b\right) $ and the equation reads 
\begin{equation}
\frac{d^{2}}{dx^{2}}h_{\left( s\right) k}\left( x\right) +\frac{1}{x}\frac{d%
}{dx}h_{\left( s\right) k}\left( x\right) +\left[ 4g\frac{T_{M}^{2}}{H^{2}}%
\frac{1}{\gamma ^{2}}-\frac{9}{4\gamma ^{2}x^{2}}\right] h_{\left( s\right)
k}\left( x\right) =0  \label{wbl}
\end{equation}
whose positive frequency solutions are again Hankel functions of the second
kind, $H_{3/2\gamma }^{\left( 2\right) }\left[ \left( 2g^{1/2}T_{M}/H\gamma
\right) u^{\gamma }\right] $. The normalized solutions to the field equation
are then 
\begin{equation}
f\left( \tau \right) =\sqrt{\frac{\pi }{2\gamma }}\frac{z^{1/2\gamma }}{%
z_{0}^{1/2\gamma }}H_{3/2\gamma }^{\left( 2\right) }\left[ z\left( \tau
\right) \right]  \label{wbn}
\end{equation}
with

\begin{equation}
z\left( \tau \right) =z_{0}\left( 1+\frac{\tau }{2}\right) ^{\gamma },\text{
\ \ \ }z_{0}=\frac{2g^{1/2}}{\gamma }\frac{T_{M}}{H}  \label{wbw}
\end{equation}

Replacing in the general expressions we get by a straightforward calculation 
\begin{equation}
\alpha _{k}=-i\sqrt{\frac{\pi }{2\gamma }}\frac{e^{ik}}{\sqrt{2k}}\left\{ %
\left[ 1+\frac{i}{k}\right] g^{1/2}\frac{T_{M}}{H}H_{3/2\gamma +1}^{\left(
1\right) }\left[ z_{0}\right] -ikH_{3/2\gamma }^{\left( 1\right) }\left[
z_{0}\right] \right\}  \label{wbr}
\end{equation}
\begin{equation}
\beta _{k}=-i\sqrt{\frac{\pi }{2\gamma }}\frac{e^{ik}}{\sqrt{2k}}\left\{
ikH_{3/2\gamma }^{\left( 2\right) }\left[ z_{0}\right] -\left[ 1+\frac{i}{k}%
\right] g^{1/2}\frac{T_{M}}{H}H_{3/2\gamma +1}^{\left( 2\right) }\left[ z_{0}%
\right] \right\}  \label{wbs}
\end{equation}
and in the limit $z_{0}\ll 1$, which is always valid, they reduce to 
\begin{equation}
\alpha _{k}=\beta _{k}\sim -i\sqrt{\frac{\gamma }{\pi }}2^{3/2\gamma
-1}\Gamma \left( \frac{2\gamma +3}{2\gamma }\right) z_{0}^{-3/2\gamma }\frac{%
e^{ik}}{k^{3/2}}  \label{hae}
\end{equation}
Observe that the apparent breakdown of these formulas at $\gamma =0$ (due to
the inadequacy of the Bessel function representation of the solution) lies
outside the physical range.

\section{Dynamics of the electromagnetic two point functions}

Cosmological particle creation is a stochastic process.\ If the particles
are charged, electromagnetic fields are induced by the created currents. But
their mean values being zero, they manifest through their variances or two
point functions. The functions we need to obtain information about the
evolution and the state of the field are respectively the {\em retarded }$%
D_{\nu \gamma }^{ret}\left( x,x^{\prime }\right) $ and the {\em Hadamard }$%
D_{1\nu \gamma }\left( x,x^{\prime }\right) $ two point functions, defined
as 
\begin{equation}
D_{\nu \gamma }^{ret}\left( x,x^{\prime }\right) =i\left\langle \left[
A_{\nu }\left( x\right) ,A_{\gamma }\left( x^{\prime }\right) \right]
\right\rangle \Theta \left( \tau -\tau ^{\prime }\right)  \label{ret}
\end{equation}
\begin{equation}
D_{1\nu \gamma }\left( x,x^{\prime }\right) =\left\langle \left\{ A_{\nu
}\left( x\right) ,A_{\gamma }\left( x^{\prime }\right) \right\} \right\rangle
\label{had}
\end{equation}

The evolution equations for this propagators, known as Schwinger-Dyson
equations (see Appendix 1), are

\begin{eqnarray}
&&\left[ \eta ^{\mu \nu }\Box _x+\left( 1-\frac 1\zeta \right) \partial ^\mu
\partial ^\nu -e^2\Gamma _{11}^{\mu \nu }\left( x,x\right) \right] D_{\nu
\gamma }^{ret}\left( x,x^{\prime }\right)  \label{abo} \\
&&+ie^2\int dx"\Sigma _{ret}^{\mu \nu }\left( x,x"\right) D_{\nu \gamma
}^{ret}\left( x",x^{\prime }\right) \left. =\right. -\delta _\gamma ^\mu
\delta \left( x-x^{\prime }\right)  \nonumber
\end{eqnarray}

\begin{eqnarray}
&&\left[ \eta ^{\mu \nu }\Box _x+\left( 1-\frac 1\zeta \right) \partial ^\mu
\partial ^\nu -e^2\Gamma _{11}^{\mu \nu }\left( x,x\right) \right] D_{1\nu
\gamma }\left( x,x^{\prime }\right)  \label{abpa} \\
&&+ie^2\int dx"\Sigma _{ret}^{\mu \nu }\left( x,x"\right) D_{1\nu \gamma
}\left( x",x^{\prime }\right) \left. =\right. -\frac{e^2}2\int dx"\Sigma
_1^{\mu \nu }\left( x,x"\right) D_{\nu \gamma }^{adv}\left( x",x^{\prime
}\right)  \nonumber
\end{eqnarray}
where $\zeta $ is a gauge fixing constant, $\Box _x=-\partial _\tau
^2+\nabla ^2$, and

\begin{equation}
\Gamma _{cd}^{\mu \nu }\left( x,x^{\prime }\right) \equiv \eta ^{\mu \nu } 
\left[ G_{cd}^{1}\left( x,x\right) +G_{cd}^{2}\left( x,x\right) \right]
\label{abia}
\end{equation}

\begin{equation}
\Sigma _{cc^{\prime },dd^{\prime }}^{\mu \nu }\equiv \eta ^{\mu \alpha
}(x)\eta ^{\nu \beta }(x")\left[ G_{cc^{\prime }}^{1}\left( x,x"\right) 
\overline{\partial }_{\alpha }\overline{\partial }_{\beta }^{\prime \prime
}G_{dd^{\prime }}^{2}\left( x,x"\right) \right]  \label{abja}
\end{equation}

$c$ and $d$ are {\em closed time path indices} whose values are $1$ for the
forward directed time path and $2$ for the backward directed time path and
the supraindex $i=1,2$ denotes the real and imaginary parts of the complex
scalar field. According to the values of the closed time path indices we
have that $G_{21}\left( x,x^{\prime }\right) \equiv \left\langle \phi \left(
x\right) \phi \left( x^{\prime }\right) \right\rangle $ is the positive
frequency two point function for the scalar field and $G_{12}\left(
x,x^{\prime }\right) \equiv \left\langle \phi \left( x^{\prime }\right) \phi
\left( x\right) \right\rangle $ the negative frequency one. With these basic
propagators we can build the antisymmetric two point function, also known as
Jordan propagator, $G\left( x,x^{\prime }\right) =G_{21}\left( x,x^{\prime
}\right) -G_{12}\left( x,x^{\prime }\right) $ and the symmetric, or Hadamard
one, $G_1\left( x,x^{\prime }\right) =G_{21}\left( x,x^{\prime }\right)
+G_{12}\left( x,x^{\prime }\right) $. $G_{11}\left( x,x^{\prime }\right)
\equiv \left\langle T\left( \phi \left( x\right) \phi \left( x^{\prime
}\right) \right) \right\rangle $ is the Feynman propagator and $G_{22}\left(
x,x^{\prime }\right) \equiv \left\langle \tilde T\left( \phi \left( x\right)
\phi \left( x^{\prime }\right) \right) \right\rangle $ the Dyson one. The
retarded and advanced two point functions are defined as $G_{ret}\left(
x,x^{\prime }\right) =iG\left( x,x^{\prime }\right) =G_{adv}\left( x^{\prime
},x\right) $, or else $G_{ret}\left( x,x^{\prime }\right) =i\left[
G_{11}\left( x,x^{\prime }\right) -G_{12}\left( x,x^{\prime }\right) \right]
$, $G_{adv}\left( x,x^{\prime }\right) =i\left[ G_{22}\left( x,x^{\prime
}\right) -G_{12}\left( x,x^{\prime }\right) \right] $. The same definitions
apply for the two point functions for the electromagnetic field, i.e. $
D_{\mu \nu }^{21}\left( x,x^{\prime }\right) \equiv \left\langle A_\mu
\left( x\right) A_\nu \left( x^{\prime }\right) \right\rangle $ is the
positive frequency two point function, $D_{\mu \nu }^{12}\left( x,x^{\prime
}\right) \equiv \left\langle A_\mu \left( x^{\prime }\right) A_\nu \left(
x\right) \right\rangle $ the negative frequency two point function, and so
on. Based on these definitions, we have built the non local kernels in eqs. 
(\ref{abo}) and (\ref{abpa}), as $\Sigma _{ret}^{\mu \nu }\left( x,x"\right)
\equiv \Sigma _{11,11}^{\mu \nu }\left( x,x"\right) -\Sigma _{12,12}^{\mu
\nu }\left( x,x"\right) \equiv \Sigma _{21,21}^{\mu \nu }\left( x,x"\right)
-\Sigma _{22,22}^{\mu \nu }\left( x,x"\right) $ and $\Sigma _1^{\mu \nu
}\left( x,x"\right) \equiv \Sigma _{12,12}^{\mu \nu }\left( x,x"\right)
+\Sigma _{21,21}^{\mu \nu }\left( x,x"\right) =\Sigma _{11,11}^{\mu \nu
}\left( x,x"\right) +\Sigma _{22,22}^{\mu \nu }\left( x,x"\right) $.

Each of equations (\ref{abo}) and (\ref{abpa}) is in fact a set of coupled
differential equations for the different components of the retarded and
Hadamard two point functions of the electromagnetic four potential. Equation
(\ref{abpa}) is the one that we shall use to evaluate the magnetic field.

\subsection{Equation for the spatial components of the electromagnetic two
point functions.}

The equations for the electromagnetic two point functions form a set, where
the equations for the pure spatial and pure temporal propagators are coupled
by the mixed functions, i.e. by two point functions with one temporal and
one spatial component. This coupling is realized through the non local
kernels, specifically by its $0-i$ components, which in general do not
vanish.

When we take into account particle creation, we see that local and non local
kernels split into a vacuum and a particle contribution, i.e. $\Gamma
_{11}^{\mu \nu }\left( x,x\right) =\Gamma _{11}^{\mu \nu \left( V\right)
}\left( x,x\right) +\Gamma _{11}^{\mu \nu \left( P\right) }\left( x,x\right) 
$ and $\Sigma _{ret\left( 1\right) }^{\mu \nu }\left( x,x"\right) =\Sigma
_{ret\left( 1\right) }^{\mu \nu \left( V\right) }\left( x,x"\right) +\Sigma
_{ret\left( 1\right) }^{\mu \nu \left( P\right) }\left( x,x"\right) $. We
are interested in the contribution of the created particles and hence shall
solve 
\begin{eqnarray}
&&\ \ \ \left[ \eta ^{\mu \nu }\Box _x+\left( 1-\frac 1\zeta \right)
\partial ^\mu \partial ^\nu -e^2\Gamma _{11}^{\mu \nu \left( P\right)
}\left( x,x\right) \right] D_{1\nu \gamma }\left( x,x^{\prime }\right)
\label{d1-ns} \\
&&\ \ +ie^2\int dx"\Sigma _{ret}^{\mu \nu \left( P\right) }\left(
x,x"\right) D_{1\nu \gamma }\left( x",x^{\prime }\right) \left. =\right. -
\frac{e^2}2\int dx"\Sigma _1^{\mu \nu \left( P\right) }\left( x,x"\right)
D_{\nu \gamma }^{adv}\left( x",x^{\prime }\right)  \nonumber
\end{eqnarray}

For our purposes of magnetic field evaluation, we shall need only the
spatial components of the propagator $D_{1lj}\left( x,x^{\prime }\right) $,
i.e. the solution to the $i-j$ equations, and in particular its transverse
parts. Since the bulk of particle creation occurs for long wavelengths, the
main contribution to the kernels comes from that sector of the spectrum (see
Appendix 2). The $0-i$ components of the non local kernels take the form of
a gradient and in that limit the mode functions depend on the momenta only
through their moduli. When transforming Fourier the spatial part of the
two point functions, there remains an integral over the momenta that vanishes
for the transverse part of the $0-i$ component. This means
that the equations for the transverse part of $D_{1lj}\left( x,x^{\prime
}\right) $ decouple from the set, a fact that facilitates enormously their
resolution. The Fourier transformed equations that we shall work with then
read (see eq. (\ref{abpa}) and Appendix 1) 
\begin{eqnarray}
&&\ \left[ \eta ^{il}\left( \partial _{\tau }^{2}+k^{2}\right) +e^{2}\int 
\frac{d^{3}p}{\left( 2\pi \right) ^{3/2}}\Gamma _{11}^{il\left( P\right)
}\left( p,\tau ,\tau \right) \right] D_{1lj}\left( k,\tau ,\tau ^{\prime
}\right)  \label{d1-eq} \\
&&\ -ie^{2}\int \frac{d^{3}p}{\left( 2\pi \right) ^{3/2}}\int d\tau "\Sigma
_{ret}^{il\left( P\right) }\left( p,k-p,\tau ,\tau "\right) D_{1lj}\left(
k,\tau ",\tau ^{\prime }\right)  \nonumber \\
&&\ \left. =\right. \frac{e^{2}}{2}\int d\tau "\Sigma _{1}^{il\left(
P\right) }\left( p,k-p,\tau ,\tau "\right) D_{lj}^{adv}\left( k,\tau
^{\prime },\tau "\right)  \nonumber
\end{eqnarray}
where $k$ is the comoving wavenumber of \ the spatially Fourier transformed
two point function of the electromagnetic field, and $p$ the corresponding
one of the scalar field. For example when replacing the mode decomposition (
\ref{haaa}) the local kernel reads 
\begin{equation}
\Gamma _{11}^{il\left( P\right) }\left( p,\tau ,\tau \right) =\eta
^{il}\left\{ 2\left| \beta _{p}\right| ^{2}\left| f_{p}\left( \tau \right)
\right| ^{2}+\alpha _{p}\beta _{p}^{\ast }f_{p}^{2}\left( \tau \right)
+\beta _{p}\alpha _{p}^{\ast }f_{p}^{\dagger 2}\left( \tau \right) \right\}
\label{aci}
\end{equation}
with $f_{p}\left( \tau \right) $ given by eq. (\ref{wbn}) or (\ref{xad}) and
the Bogolyubov coefficients by (\ref{hae}), and similarly the non local
kernels. We assume that the vacuum part may be absorbed into a
renormalization of the classical action, the remainder being negligible.

\subsection{Computing the kernels}

In this section we shall compute the kernels found in the equations (\ref
{d1-eq}) for the Maxwell field propagators. It may be checked that the
contribution from the large wavenumber sector is negligible (see Appendix 2)
in comparison to the infrared one, a fact that can be understood by looking
at the expressions for the Bogolyubov coefficients, eqs. (\ref{xae}), (\ref
{xaf}) and (\ref{hae}).{\em \ }The mode functions are given by eq. (\ref{wbn}
) and we see that they do not depend on the wavenumbers $p$. The Bogolyubov
coefficient on the other side retain their full momentum dependence.

When we replace for the modes and coefficients, eqs. (\ref{wbn}) and (\ref
{hae}), we find a logarithmic divergence in the momentum integral of the
local kernel eq. (\ref{aci}). We introduce an infrared cut-off, which we can
choose as the comoving, dimensionless mode $\Upsilon $ corresponding to the
original inflationary patch that gave rise to our universe. For the upper
limit we take $\Delta \sim T_{M}/H$. We then have 
\begin{eqnarray}
\Gamma _{11}^{il}\left( \tau ,\tau \right) &=&\int_{\Upsilon }^{\Delta }%
\frac{d^{3}p}{\left( 2\pi \right) ^{3/2}}\Gamma _{11}^{il\left( P\right)
}\left( p,\tau ,\tau \right) \simeq \eta ^{il}\tilde{\Gamma},  \label{acna}
\\
\tilde{\Gamma} &=&2^{3/\gamma -1}\Gamma ^{2}\left( \frac{2\gamma +3}{2\gamma 
}\right) \ln \left( \frac{\Delta }{\Upsilon }\right) z_{0}^{-4/\gamma
}F^{2}\left( z\right)  \label{acnb}
\end{eqnarray}
where

\begin{equation}
F\left( z\right) =z^{1/2\gamma }J_{3/2\gamma }\left( z\right)  \label{acnc}
\end{equation}
$J_{3/2\gamma }\left( z\right) $ being a Bessel function of the first kind
an $z$ and $z_0$ being defined above in eq. (\ref{wbw}).

The non local noise kernel has no infrared divergences and is $\Delta ^2$
times smaller than the local one, with $\Delta \sim T_M/H\ll 1$ (see
Appendix 2). This difference in the orders of magnitude will never decrease,
(even if we evaluate the remaining time integral in the non local part of
the equation) and therefore for our purposes we can ignore the contribution
of this kernel.

This is a very important approximation, not only from the technical point of
view but also because it shows that dissipation due to the 
ordinary conductivity of the own charges
is negligible. 

We now turn to the Hadamard noise kernel\ $\Sigma _{1}^{ij\left( N\right)
}\left( p,k-p,\tau ,\tau "\right) $ that is the source in the equation (\ref
{d1-eq}) for $D_{1}^{\left( N\right) }\left( k,\tau 
,\tau ^{\prime }\right) $. The same considerations made for the retarded 
kernel about the gradient
structure of the $0-i$ component of the two point function hold for this
one. So we shall only need to evaluate the pure spatial components, given by 
\begin{eqnarray}
\Sigma _{1}^{il\left( P\right) }\left( p,k-p,\tau ,\tau "\right) &\sim &%
\frac{p^{i}p^{l}}{p^3\left| \bar{k}-\bar{p}\right|^3 }S_{1}\left( \tau
,\tau ^{\prime \prime }\right) ,  \label{acnd} \\
S_{1}\left( \tau ,\tau ^{\prime \prime }\right) &=&2^{6/\gamma -1}\Gamma
^{4}\left( \frac{2\gamma +3}{2\gamma }\right) z_{0}^{-8/\gamma }F^{2}\left[
z\left( \tau \right) \right] F^{2}\left[ z\left( \tau "\right) \right]
\label{acne}
\end{eqnarray}
It is remarkable that the same function $F$ appears in this kernel and in (%
\ref{acnb}).

We see that after replacing the modes and Bogolyubov coefficients the
resulting kernels given by eq. (\ref{acnb}) and (\ref{acne}) are real
functions. This happens because of exp. (\ref{hae}) for the Bogolyubov
coefficients and the fact that the modes (\ref{wbn}) do not depend on the
wavenumbers. In other words, after replacing the Bogolyubov coefficients,
the modes (\ref{wbn}) combine to give real functions that oscillate
coherently, an indication that they are indeed superhorizoned and hence
frozen.

\subsection{Evolution equations for the magnetic field}

After all the considerations we have made, the equation for the transverse
part of the pure spatial Hadamard two point function for the electromagnetic
four potential reads 
\begin{equation}
\left[ \eta ^{il}\left( \partial _\tau ^2+k^2\right) +e^2\Gamma
_{11}^{il\left( P\right) }\left( \tau ,\tau \right) \right] D_{1lj}\left(
k,\tau ,\tau ^{\prime }\right) =\Xi _{ij}\left( k,\tau ,\tau ^{\prime
}\right)  \label{adn}
\end{equation}
with{\em \ } 
\begin{equation}
\Xi _{ij}\left( k,\tau ,\tau ^{\prime }\right) =\frac{e^2}2\int_0^{\tau
^{\prime }}d\tau "\int_\Upsilon ^\Delta \frac{d^3p}{\left( 2\pi \right)
^{3/2}}\Sigma _1^{il}\left( p,k-p,\tau ,\tau "\right) D_{lj}^{ret}\left(
k,\tau ^{\prime },\tau "\right)  \label{ado}
\end{equation}

Let us introduce the Hadamard two point function for the magnetic field: 
\begin{equation}
\left\langle \left\{ B_{i}\left( \tau ,\bar{r}\right) ,B_{j}\left( \tau
^{\prime },\bar{r}^{\prime }\right) \right\} \right\rangle =H^{4}\epsilon
^{ikl}\epsilon ^{jk^{\prime }l^{\prime }}\int \frac{d^{3}k}{\left( 2\pi
\right) ^{3/2}}k_{k}k_{k^{\prime }}D_{1ll^{\prime }}\left( k,\tau ,\tau
^{\prime }\right) \exp \left[ i\bar{k}.\left( \bar{r}-\bar{r}^{\prime
}\right) \right]  \label{adq}
\end{equation}
where the prime refers to the prime coordinate and $H^{4}$ gives the
dimensions.

We are interested in the field $B_{i}\left( \tau ,\lambda \right) $ coherent
over a scale $\lambda $, so we must filter the high frequency contribution
with a window function of ''size'' $\lambda $, i.e. we must evaluate 
\begin{equation}
\frac{1}{V_{\lambda }^{2}}\int_{V_{\lambda }}d^{3}r\int_{V_{\lambda
}}d^{3}r^{\prime }\left\langle \left\{ B_{i}\left( \tau ,\bar{r}\right)
,B_{j}\left( \tau ^{\prime },\bar{r}^{\prime }\right) \right\} \right\rangle
\label{adu}
\end{equation}
where $V_{\lambda }\sim \lambda ^{3}$ is a comoving volume in which we seek
homogeneity. Eq. (\ref{adu}) amounts to calculate 
\begin{equation}
\frac{1}{V_{\lambda }}\int_{V_{\lambda }}d^{3}r\exp \left[ i\bar{k}.\bar{r}%
\right] =\frac{1}{V_{\lambda }}\int_{V_{\lambda }}d^{3}r^{\prime }\exp \left[
-i\bar{k}.\bar{r}^{\prime }\right] \equiv W_{\lambda }\left( k\right)
\label{adv}
\end{equation}
This window function can be approximated by 
\begin{eqnarray}
W_{\lambda }\left( k\right) &\sim &1\quad if\quad k\leq K=1/\lambda
\label{adw} \\
&=&0\quad \text{otherwise}  \nonumber
\end{eqnarray}
which can be implemented as a cut-off in the $k$-momentum integral.

For our purposes it is enough to compute the self-correlation $\left\langle
\left\{ B^{i}\left( \tau ,\lambda \right) ,B_{i}\left( \tau ^{\prime
},\lambda \right) \right\} \right\rangle .$ From the equation for the
Hadamard propagator we get

\begin{eqnarray}
&&\left[ \frac{d^{2}}{d\tau ^{2}}+e^{2}\Gamma \left( \tau ,\tau \right) %
\right] \left\langle \left\{ B^{i}\left( \tau ,\lambda \right) ,B_{i}\left(
\tau ^{\prime },\lambda \right) \right\} \right\rangle \left. =\right.
\label{adx} \\
&&2e^{2}H^{4}\int_{0}^{K}\frac{d^{3}k}{\left( 2\pi \right) ^{3/2}}%
\int_{\Upsilon }^{\Delta }\frac{d^{3}p}{\left( 2\pi \right) ^{3/2}}%
\int_{0}^{\tau }d\tau ^{\prime \prime }\frac{\left| \bar{k}\times \bar{p}%
\right| ^{2}}{p^{3}\left| k-p\right| ^{3}}S_{1}\left( \tau ^{\prime \prime
\prime },\tau ^{\prime \prime }\right) D^{ret}\left( \tau ^{\prime },\tau
^{\prime \prime }\right)  \nonumber
\end{eqnarray}
where we used the fact that, because $K\ll 1$, the spatial gradients are
negligible, and $D^{ret}$ is therefore independent of the wavenumber. \ The
momentum integral can be easily evaluated and we obtain

\begin{equation}
\left[ \frac{d^{2}}{d\tau ^{2}}+e^{2}\Gamma \left( \tau ,\tau \right) \right]
\left\langle \left\{ B^{i}\left( \tau ,\lambda \right) ,B_{i}\left( \tau
^{\prime },\lambda \right) \right\} \right\rangle
=2e^{2}H^{4}K^{4}\int_{0}^{\tau }d\tau ^{\prime \prime }S_{1}\left( \tau
,\tau ^{\prime \prime }\right) D^{ret}\left( \tau ^{\prime },\tau ^{\prime
\prime }\right)  \label{ady}
\end{equation}

\section{Equivalent stochastic problem}

Rather than solving directly equation (\ref{ady}), we move to a physically
more transparent language and translate our problem to a stochastic
formulation. We introduce a stochastic field $B\left( \tau \right) $ obeying
the Langevin equation

\begin{equation}
\left[ \frac{d^{2}}{d\tau ^{2}}+e^{2}\Gamma \left( \tau ,\tau \right) \right]
B\left( \tau \right) =\xi \left( \tau \right)  \label{paa}
\end{equation}
where $\xi \left( \tau \right) $ is Gaussian noise with zero mean and self
correlation 
\begin{equation}
\left\langle \xi \left( \tau \right) \xi \left( \tau ^{\prime \prime
}\right) \right\rangle _{\xi }=4e^{2}H^{4}K^{4}S_{1}\left( \tau ,\tau
^{\prime \prime }\right)  \label{paaa}
\end{equation}
It is then easy to see that

\begin{equation}
\left\langle \left\{ B^{i}\left( \tau ,\lambda \right) ,B_{i}\left( \tau
^{\prime },\lambda \right) \right\} \right\rangle =\left\langle B\left( \tau
\right) B_{i}\left( \tau ^{\prime }\right) \right\rangle _{\xi }  \label{pab}
\end{equation}
Moreover, because of the structure of the self-correlation, we may realize
the noise as

\begin{equation}
\xi \left( \tau \right) =eH^{2}K^{2}2^{3/\gamma }\Gamma ^{2}\left( \frac{
2\gamma +3}{2\gamma }\right) z_{0}^{-4/\gamma }F^{2}\left[ z\left( \tau
\right) \right] X  \label{pac}
\end{equation}
where $X$ is a single gaussian variable with $\left\langle X\right\rangle =0$
and $\left\langle X^{2}\right\rangle =1$.

Since the equation is linear, we may write

\begin{equation}
B\left( \tau \right) =eH^{2}K^{2}2^{3/\gamma }\Gamma ^{2}\left( \frac{
2\gamma +3}{2\gamma }\right) z_{0}^{-4/\gamma }XB_{s}\left( \tau \right) 
\label{pad}
\end{equation}
The equation to be solved is then

\begin{equation}
\left[ \frac{d^{2}}{d\tau ^{2}}+C^{2}F^{2}\left[ z\left( \tau \right) \right]
\right] B_{s}\left( \tau \right) =F^{2}\left[ z\left( \tau \right) \right]
\label{pae}
\end{equation}
where

\begin{equation}
C^{2}=\tilde e^{2}z_{0}^{-4/\gamma }  \label{paf}
\end{equation}
where 
$e^22^{3/\gamma -1}\Gamma
^{2}\left[ \left( 2\gamma +3\right) /2\gamma \right] \ln \left( \Delta
/\Upsilon \right) =\tilde e^2$ with the boundary conditions $B_{s}
\left( 0\right) =\dot{B}_{s}\left( 0\right) =0$. 
Eq. (\ref{pae}) resembles the London
equation in a superconducting medium \cite{schrieffer}. This means that the
induced field will be weaker than the obtained in the case of free
propagation. We may write $B_{s}\left( \tau \right) =C^{-2}\left(
1+B_{h}\left( \tau \right) \right) $ and the equation for $B_{h}\left( \tau
\right) $ reads

\begin{equation}
\left[ \frac{d^{2}}{d\tau ^{2}}+C^{2}F^{2}\left[ z\left( \tau \right) \right]
\right] B_{h}\left( \tau \right) =0  \label{pag}
\end{equation}
with initial conditions $B_{h}\left( 0\right) =-1,$ $\dot{B}_{h}\left(
0\right) =0$. It is convenient to adopt the variable $z$ as independent
variable. Equation (\ref{pag}) now reads

\begin{equation}
\left[ \frac{d^{2}}{dz^{2}}+\frac{\left( \gamma -1\right) }{\gamma z}
\frac{d}{dz}+c^{2}g^{2}\left( z\right) \right] B_{h}\left( z\right) =0  
\label{pah}
\end{equation}
where $c^{2}=4\tilde e^{2}z_{0}^{-6/\gamma }/\gamma ^{2}$ and 
$g^{2}\left( z\right)
=z^{-2\left( \gamma -1\right) /\gamma }F^{2}\left[ z\right] =z^{\left(
3-2\gamma \right) /\gamma }J_{3/2\gamma }^{2}\left( z\right) $. The magnetic
field $B\left( \tau \right) $ will then be given by
\begin{equation}
B\left( \tau \right) \sim H^{2}K^{2}\left( 1+B_{h}\left( \tau \right)
\right)   \label{pagg}
\end{equation}

Because of the oscillatory behavior of $g^{2}\left( z\right) $ in equation 
(\ref{pah}) we expect that the attenuation of the induced magnetic field due 
to the London currents will not be so strong. Besides, due to the form
of $g^2\left( z\right) $ we might expect some sort of resonant growth. This
can be understood as follows.

Defining 
\begin{equation}
B_{h}\left( z\right) =z^{-\left( \gamma -1\right) /2\gamma }b_{h}\left(
z\right) \label{xxa}
\end{equation}
and replacing in eq. (\ref{pah}) we get 

\begin{equation}
\ddot{b}_{h}\left( z\right) +\left[ \frac{\gamma ^{2}-1}{4\gamma ^{2}z^{2}}+
\frac{4\tilde{e}^{2}}{\gamma ^{2}z_{0}^{6/\gamma }}\frac{J_{3/2\gamma
}^{2}\left( z\right) }{z^{\left( 2\gamma -3\right) /\gamma }}\right]
b_{h}\left( z\right) =0 \label{xxb}
\end{equation}
As in general it is satisfied that $4\tilde{e}^{2}z_{0}^{-6/\gamma }\gg
\left( \gamma ^{2}-1\right) /4$ we can approximate equation (\ref{xxb}) by

\begin{equation}
\ddot{b}_{h}\left( z\right) +c^{2}
\frac{J_{3/2\gamma }^{2}\left( z\right) }{
z^{\left( 2\gamma -3\right) /\gamma }}b_{h}\left( z\right) =0
\label{xxc}
\end{equation}

Consider the asymptotic form of the Bessel functions, for $z \gg 1$, i.e.
$J_{\nu }\left( z \right) \sim 
\sqrt{2/\left( \pi z\right) }\cos \left[ z-\nu \pi /2-\pi /4
\right] $. The second term in eq. (\ref{xxc}) can be then written as
$q\left( \gamma,T_M/H, z\right) \cos^2 \left[ z-\left( 3/2\gamma\right) \pi 
/2-\pi /4 \right] $ with

\begin{equation}
q\left( \gamma,T_M/H, z\right) \sim 
\frac{4\tilde e^2}{\gamma^2 z_0^{6/\gamma }}
z^{3\left( 1-\gamma\right)/\gamma}\label{xxd}
\end{equation}
For $q\left( z\right) \gg 1$, or equivalently until

\begin{equation}
z\approx \left(\frac{4\tilde e^2}{\gamma^2 z_0^{6/\gamma}}  \right)
^{\gamma /\left(3\gamma -3\right) }\label{xxe}
\end{equation}
the growth of the field
intensity could be described in terms of stochastic resonance (originally 
developed in Ref. \cite{fujisaki} and extended in Ref. \cite{kofman}),
because of violation of the adiabatic condition.

In order to obtain an accurate estimation of the magnetic field, 
we shall integrate eq. (\ref{pah}) numerically and evaluate the induced
field from eq. (\ref{pagg}). This procedure will also allow us to
confirm the previous heuristic analysis of stochastic resonance.
In view of the uncertainties in the actual
values of $\gamma $ and $T_{M}$ we shall not atempt to estimate the induced
field for a wide interval of their possible values, but shall illustrate the
effect by considering one specific choice, namely $\gamma =5/2$ and 
$T_{M}/H=10^{-2}$. We also fix $\tilde e^2 =1$ for all possible values
of $\gamma$ (in fact $\tilde e^2$ is a slowly varying function of 
$\gamma$ and this choice does not affect the outcomes of the numerical
integration). This value of $\gamma $ corresponds to $b=1/4$ and
describes a strong deviation of the thermal mass from conformal thermal
evolution. We recall that the parameter $b$ does not describe the evolution
of the temperature of the reheating plasma, i.e. the one formed by the
inflaton decays, but the one of the thermal shift of the scalar charges. 
For this values of the physical parameters, according to eq.
(\ref{xxe}) we might expect resonant growth until $z \approx 500$.

Considering the maximum allowed value for the Hubble constant during
inflation, namely $H=10^{13}$ GeV, $T_{M}/H=10^{-2}$ corresponds to 
$T_{M}=10^{11}$ GeV.

\subsection{Limits of validity}

Eq. (\ref{pah}) correctly describes the evolution of $B_{h}$ during a time
interval in which the equations obtained in the leading order of the $1/N$
approximation, eqs. (\ref{abo}) and (\ref{abpa}) are valid. To estimate this
interval we can compare the growth of the electromagnetic energy density
with the one of the scalar field: assuming efficient conversion of scalar
energy into electromagnetic one, the moment at which the latter overtakes
the former can be considered as the limiting time to integrate the
equations. In fact, in the leading order of the {\em 1/N} approximation the
equation for the scalar field propagator shows no back-reaction from the
magnetic field (see Appendix 1). This means that to this
order the induced field is very weak and hence its energy density must be
smaller than the one of the scalar field. When this condition is broken, 
the evolution equations for the two point functions cease
to be valid. The energy density is given by the $00$ component of the stress
energy tensor of the system of fields, $T^{00}$. 
To evaluate this tensor component we consider the
electromagnetic field as classical and the scalar one as quantum. We can
split $T^{00}$ in a pure scalar contribution and an electromagnetic part as 
\begin{equation}
T_{\varphi }^{00}=H^{4}\left[ \dot{\varphi}^{2}-2\frac{\dot{a}}{a}\varphi 
\dot{\varphi}+\frac{1}{2}\eta ^{ij}\partial _{i}\varphi \partial _{j}\varphi
+\left( \frac{m^{2}\left( \tau \right) }{H^{2}}a^2\left( \tau\right)
+\frac{\dot{a}^{2}}{a^{2}
}\right) \varphi ^{2}\right]   \label{taa}
\end{equation}
\begin{eqnarray}
T_{A}^{00} &=&H^{4}\frac{1}{4}\left[ \tilde{F}^{ij}\tilde{F}_{ij}+2\tilde{F}
^{0j}\tilde{F}_{0j}\right]   \label{tab} \\
&&+H^{4}\left[ \eta ^{ij}eA_{i}\left( \partial _{j}\varphi _{1}\varphi
_{2}-\varphi _{1}\partial _{j}\varphi _{2}\right) +\frac{1}{2}\eta
^{ij}e^{2}A_{i}A_{j}\left( \varphi _{1}^{2}+\varphi _{2}^{2}\right) \right] 
\nonumber
\end{eqnarray}
where it is understood that all quantities between brackets are
dimensionless. We shall compare $\left\langle T_{\varphi }^{00}\right\rangle 
$ with $\left\langle T_{A}^{00}\right\rangle $: the instant at which the the
latter equals the former will be considered as the time at which the
equations obtained in the large $N$ limit cease to be valid.

\subsubsection{Stress energy tensor for the scalar field}

Considering the electromagnetic field as classical and the scalar as
quantum, and defining $T^{00}=H^{4}\tilde{T}^{00}$we can write 
\begin{eqnarray}
\left\langle \tilde{T}_{\varphi }^{00}\right\rangle &=&\int \frac{d^{3}p}{
\left( 2\pi \right) ^{3/2}}\left\{ \dot{\varphi}_{p}^{\ast }\left( \tau
\right) \dot{\varphi}_{p}\left( \tau \right) -\frac{\dot{a}}{a}\left[
\varphi _{p}^{\ast }\left( \tau \right) \dot{\varphi}_{p}\left( \tau \right)
+\dot{\varphi}_{p}^{\ast }\left( \tau \right) \varphi _{p}\left( \tau
\right) \right] \right.  \label{tad} \\
&&+\left. \eta ^{ij}p_{i}p_{j}\varphi _{p}^{\ast }\left( \tau \right)
\varphi _{p}\left( \tau \right) +\left( a^{2}\frac{m^{2}\left( \tau \right) 
}{H^{2}}+\frac{\dot{a}^{2}}{a^{2}}\right) \varphi _{p}^{\ast }\left( \tau
\right) \varphi _{p}\left( \tau \right) \right\}  \nonumber
\end{eqnarray}
Writing the modes as $\varphi _{p}\left( \tau \right) \sim \alpha
_{p}f_{p}\left( \tau \right) +\beta _{p}f_{p}^{\ast }\left( \tau \right) $
with the modes $f_{p}\left( \tau \right) $ given by eq. (\ref{wbn}) and the
Bogolyubov coefficients by eq. (\ref{hae}), the contribution from the
created particles to $\left\langle \tilde{T}_{\varphi }^{00}\right\rangle $
reads 
\begin{eqnarray}
\left\langle \tilde{T}_{\varphi }^{00}\right\rangle _{\beta } &\simeq
&\int_{\Upsilon }^{\Delta }\frac{d^{3}p}{\left( 2\pi \right) ^{3/2}}\left|
\beta _{p}\right| ^{2}\left\{ \left[ \dot{f}\left( \tau \right) +\dot{f}
^{\ast }\left( \tau \right) \right] ^{2}-2\frac{\dot{a}}{a}\left[ \dot{f}
\left( \tau \right) +\dot{f}^{\ast }\left( \tau \right) \right] \left[
f\left( \tau \right) +f^{\ast }\left( \tau \right) \right] \right.
\label{taf} \\
&&+\left. \eta ^{ij}p_{i}p_{j}\left[ f\left( \tau \right) +f^{\ast }\left(
\tau \right) \right] ^{2}+\left( a^{2}
\frac{m^{2}\left( \tau \right) }{H^{2}}
+\frac{\dot{a}^{2}}{a^{2}}\right) \left[ f\left( \tau \right) +f^{\ast
}\left( \tau \right) \right] ^{2}\right\}  \nonumber
\end{eqnarray}
Neglecting the gradient term because it will give a negligible contribution
in comparison to the other terms and performing the momentum integral, we
get 
\begin{eqnarray}
\left\langle \tilde{T}_{\varphi }^{00}\right\rangle _{\beta } &=&\frac{4\pi 
}{\left( 2\pi \right) ^{3/2}}\frac{\gamma }{\pi }z_{0}^{-3/\gamma }\times
\left\{ a^{\gamma -1}\frac{T_{M}^{2}}{H^{2}}\left[ f\left( \tau \right)
+f^{\ast }\left( \tau \right) \right] ^{2}\right. \\
&&+\left. \left( \left[ \dot{f}\left( \tau \right) +\dot{f}^{\ast }\left(
\tau \right) \right] -\frac{\dot{a}}{a}\left[ f\left( \tau \right) +f^{\ast
}\left( \tau \right) \right] \right) ^{2}\right\}  \nonumber
\end{eqnarray}
where we have used $a^{2}m^{2}\left( \tau \right) /H^{2}=a^{\gamma
-1}T_{M}/H $.

Replacing the modes in terms of the Bessel functions through eq. 
(\ref{wbn}), the scale factor and its derivative in terms of $z$ and $z_{0}$ 
and using 
$T_{M}/H=\gamma z_{0}/2$ we are left with 
\begin{equation}
\left\langle \tilde{T}_{\varphi }^{00}\right\rangle =
\frac{2\pi \gamma ^{2}}{
\left( 2\pi \right) ^{3/2}}\frac{z^{2-1/\gamma }}{z_{0}^{2/\gamma }}\left\{
J_{3/2\gamma }^{2}\left[ z\left( \tau \right) \right] \ +
J_{3/2\gamma +1}^{2}
\left[ z\left( \tau \right) \right] \right\}  \label{tah}
\end{equation}
For $z\gg 1$ we can replace the asymptotic expressions of the Bessel
functions \cite{abramowitz}, i.e. $J_{\nu }\left[ z\left( \tau \right) 
\right] \sim \sqrt{2/\left( \pi z\right) }\cos \left[ z-\nu \pi /2-\pi /4
\right] $ and so obtain the final expression for $\left\langle \tilde{T}
_{\varphi }^{00}\right\rangle $: 
\begin{equation}
\left\langle \tilde{T}_{\varphi }^{00}\right\rangle \sim 
\frac{4\gamma ^{2}}{
\left( 2\pi \right) ^{3/2}}\frac{z^{1-1/\gamma }}{z_{0}^{2/\gamma }}
\label{tak}
\end{equation}
which is plotted in Fig. 3.
\subsubsection{Stress-energy tensor of the electromagnetic field}

As stated in eq. (\ref{tab}), the stress energy tensor for the
electromagnetic field consists of the pure electromagnetic and the
interaction energy density. Proceding with eq. (\ref{tab}) similarly as for $
T^{00}$ we are left with 
\begin{equation}
\left\langle \tilde{T}_{A}^{00}\right\rangle =\frac{1}{2}\left( \tilde{B}
^{2}+\tilde{E}^{2}\right) +e^{2}\frac{B^{2}}{K^{2}}\frac{z^{1/\gamma }}{
z_{0}^{4/\gamma }}J_{3/2\gamma }^{2}\left( z\right)   \label{tal}
\end{equation}
Using that $\tilde{B}=B/H^{2}$ and 
\begin{equation}
\tilde{E}=\frac{d}{d\tau }\frac{\tilde{B}}{K}=\frac{\gamma }{2K}\frac{
z_{0}^{1/\gamma }}{z^{1/\gamma -1}}\frac{d\tilde{B}}{dz}  \label{tan}
\end{equation}
we get  
\begin{equation}
\left\langle \tilde{T}_{A}^{00}\right\rangle =\frac{1}{2}\left[ \tilde{B}
^{2}+\frac{\gamma ^{2}}{4K^{2}}\frac{z_{0}^{2/\gamma }}{z^{2/\gamma -2}}
\left( \frac{d\tilde{B}}{dz}\right) ^{2}\right] +e^{2}\frac{\tilde{B}^{2}}{
K^{2}}\frac{z^{1/\gamma }}{z_{0}^{4/\gamma }}J_{3/2\gamma }^{2}\left(
z\right)   \label{tao}
\end{equation}
We solved numerically eq. (\ref{pah}) and with the outcomes reconstruct 
$B$ from eq. (\ref{pagg}). Eq. (\ref{pagg}) gives us the field coherent over
a comoving scale $K^{-1}$ in which we might be interested. We see from that
equation that the field intensity grows as $K^{2}$. In order to make a
honest computation of $\left\langle \tilde{T}_{A}^{00}\right\rangle $ we
must take into account the highest possible magnetic field intensity produced
by our mechanism, and this is produced at a scale $K_{\max }\sim T_{M}/H$,
i.e. the cut-off that we used to split the momentum spectrum. In Fig. 3 we
have also plotted the outcomes of these calculations.

\section{Magnetic Field Evaluation}

To estimate the intensity of the induced magnetic field, coherent over a
physical scale $\kappa ^{-1}$ at a given time $\tau $, we must replace 
\begin{equation}
a\left( \tau \right) \kappa =HK  \label{pao}
\end{equation}
So, from eqs. (\ref{pagg}) we get 
\begin{equation}
B_{phys}\left( \tau \right) =\frac{B\left( \tau \right) }{a^{2}\left( \tau
\right) }\sim \kappa ^{2}\left( 1+B_{h}\left( \tau \right) \right) 
\label{pap}
\end{equation}

With the parameters given above, i.e. $\gamma =5/2$ and $T_{M}/H=10^{-2}$
and fixing $\tilde e^2 =1$ for all values of $\gamma$,
we have numerically solved eq. (\ref{pah}) and show the results in Fig. 1.
We see that $B_{h}\left[ \left( z\right) \right] $ oscillates with an
amplitude that increases exponentially.  We have estimated numerically the
exponent of the exponential envelope in the $z$ interval considered, finding 
$\mu =2/15$. The oscillatory shape of the curve clearly suggests that the 
growth of the field is non-adiabatic, and hence we can expect stochastic 
amplification for all values of the physical parameters and during a
time interval such that $q\left( \gamma,T_M/H, z\right)\gg 1$.
 
\epsfxsize 4in
\begin{figure}[htpb]
\centering\leavevmode
\epsfbox{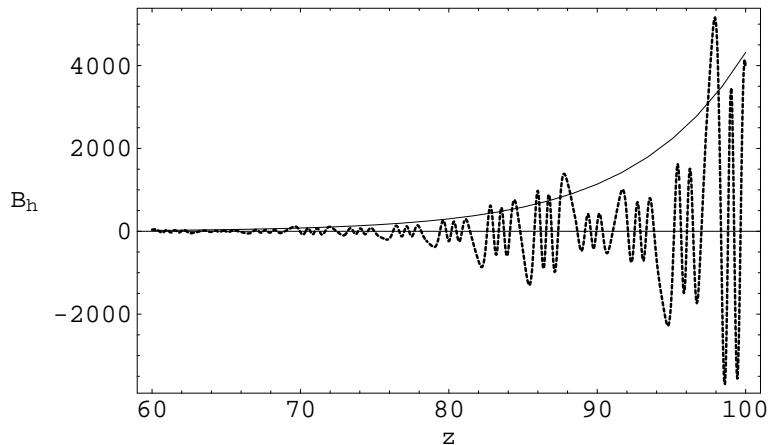}
\caption{Plot of $B_{h}$ vs. $z$: the magnetic field intensity grows
exponentially, as is shown by the dashed line. The numerically estimated 
exponent is $\mu = 2/15$ and the amplitude $B_{0h}=300$}
\label{$B_h$ vs $z$}
\end{figure}

The fact that the field grows non adiabatically can be seen in Fig. 2,
where we have plotted $B_h$ and $g(z)$ vs $z$. We can clearly seen that
the maximum growth of the field occurs when $g(z)=0$.

\epsfxsize 4in
\begin{figure}[htpb]
\centering\leavevmode
\epsfbox{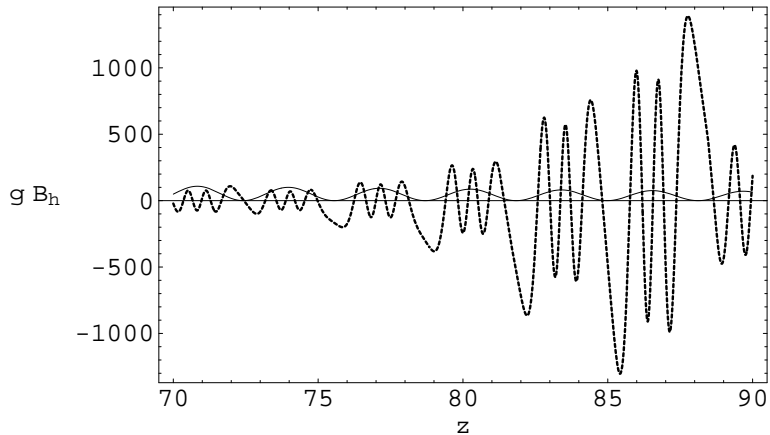}
\caption{Plot of $B_{h}$ and $g(z)$ vs. $z$: The dashed line corresponds
to $B_h$, as in Fig. 1, while the filled one to $g(z)$. We see that $B_h$ 
grows appreciably when $g(z)=0$, a fact that confirms the non adiabaticity
of the induction process.}
\label{$B_h$ and $g(z)$ vs $z$}
\end{figure}

To calculate $\left\langle T_{A}^{00}\right\rangle $ we considered the
exponential envelope of the numerical evaluation and its derivative
to estimate the magnetic and electric fields respectively.
It can be seen from Fig. 3 that at $z\sim 95$
the magnetic energy density equals the one of the scalar field and hence our
equations cease to be valid. At this time, $\left( 1+B_{h}\left( \tau \right)
\right) \sim 10^{3}$. The growth of the field can therefore be understood
in terms of stochastic resonance since, as stated in the previous subsection,
we can expect this behavior for $z_0 \leq z < 500$, and the allowed time 
interval of integration of our equations is well inside this one.

\epsfxsize 4in
\begin{figure}[htpb]
\centering\leavevmode
\epsfbox{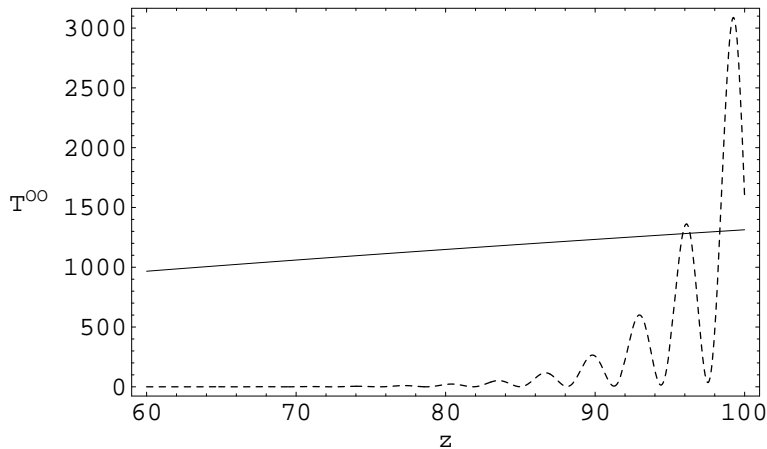}
\caption{Plot of the energy density of the scalar field (solid line) and of the
electromagnetic field (dashed, oscillatory curve) as a function of $z$. The 
magnetic energy overtakes the scalar one at $z\sim 95$}
\label{$T^{00}$ vs $z$}
\end{figure}

For example, if we consider the comoving galactic scale today, 
$\kappa \simeq 10^{-38}$
GeV and use the equivalence $1GeV^{2}\simeq 10^{20}$ Gauss, we obtain a
magnetic field intensity $B_{phys}^{gal}\left( \tau _{tod}\right) \simeq
10^{53}$ Gauss, a value too weak to seed the galactic dynamo.

\section{Discussion}

In this paper we have evaluated in a self-consistent manner the induction of
magnetic fields from electric currents created at the inflation-reheating 
transition. As the mean value of the current vanishes, this current arises 
due to the quantum and stochastic fluctuations around the mean.

Using techniques specially suited for studying quantum fields out of
equilibrium such as the 2PI closed time path effective action together with
the large {\em N} approximation, we have obtained consistent evolution
equations for the two point functions for the scalar and electromagnetic
fields, known as Schwinger - Dyson equations.

To evaluate the field we took into account the possible couplings of the
created particles to the other fields present in the reheating epoch, by
using a phenomenological expression of a temperature dependent mass. Any
other coupling to the forming matter fields could be account as an
external conductivity, which would give ordinary dissipation already
discussed in the literature \cite{ale1,misha1,ale2}.

The first important result of our work is that at leading order in the 
{\em 1/N} 
expansion there is no back-reaction of the induced magnetic field on its 
sources, and hence the charged field propagates as a free field.

When computing particle creation, we find that the largest number is created 
in the infrared portion of the spectrum, giving this sector
the main contribution to the different terms that enter in the Schwinger
Dyson equations.

The evolution equations for the magnetic field two point function show
two kernels, a local and a non local one. When we evaluate these kernels with
the infrared modes and Bogolyubov coefficients, we find that the local kernel
dominates over the non local, dissipative one, by several orders of magnitude 
during all reheating. Physically this means that dissipation in our system
of fields will not be due to an ordinary electric conductivity.
This fact also allows the 
equations for the pure spatial components of the electromagnetic two point
functions to decouple, a fact that facilitates enormously their resolution.

In order to estimate the induced field we translate our equations to a more
familiar and physically clearer Langevin equation, which due to the presence of
the local kernel,
resembles the London equation for a superconducting medium. This means that
there exists attenuation produced
by the same created particles, which is not due to ordinary conductivity but 
to screening due to the
fact that the created particles behave as a superconducting medium. This is
the second important result of our paper and could be obtained because
we work with the full quantum field theoretical features of our system.

To visualize the process of field induction, we have solved numerically 
eq. (\ref{pah}) for $\gamma = 5/2$ and $T_M/H=10^{-2}$
and found that, due to the fact that the London current oscillates, 
the damping is not perfect. The magnetic field grows exponentially, 
and the growth rate in the analyzed case is  $\mu \simeq 2/15$.
This exponential growth is non adiabatic and generic for all superhorizoned 
modes of the magnetic field, and could be cast in the framework of
stochastic resonance. In view of our ignorance about the
actual couplings of our scalar field and specially about the 
very process of reheating, we did not pursue a thorough analysis of the 
possible parameter space in order to give bounds to the final field intensity.

In general we can estimate the upper limit of integration of eq. (\ref{pah}) 
as the one at 
which the energy density in the electromagnetic field overtakes the one in
the scalar field. This estimation is based on the fact that to leading order
in the $1/N$ expansion, there is no back-reaction of the electromagnetic fields
on their sources. When this condition is violated, i.e. when the 
electromagnetic field becomes so intense that its energy density equals the
one of its sources, the approximation breaks down and the equations cease to
be valid. 

For the set of parameters used in this paper to illustrate the field induction,
this time interval is rather short and the resulting field intensity
too weak to have an astrophysical effect.

Due to the different powers of the momenta, the kinetic
term in eq. (\ref{tao}) for the $00$ component of the electromagnetic
stress-energy tensor is negligible with respect to the interaction term, i.e.
$e^2A^2\langle \varphi^2\rangle$. The growth of this term is given by the 
growth in the magnetic field, which in turn is determined by
the index of the exponential. 
Therefore the smaller this index the later
the magnetic energy will catch up with the one of the scalar field, which means
that the final field intensity will be higher.

D. Boyanovsky et al. have shown in Ref. \cite{boya-cond} that the
electrical conductivity of a primordial plasma might grow exponentially.
Therefore
it could happen that outside the range of validity of the $1/N$ approximation,
the conductivity might damp the amplification of the field.
The final answer, however, can only be given from the analysis of the
higher order corrections to the magnetic field evolution.

Given our present ignorance about the actual values of the physical 
parameters
that determine that index, a very small index and hence a stronger 
field cannot be ruled out. Also the obtained intensities could be
preamplified by some primordial process, as for example the one 
considered in Ref. \cite{finelli,tsujikawa}, and then lead to acceptable 
values to seed the subsequent amplifying mechanisms.

\begin{center}
\-{\bf Acknowledgements}
\end{center}

We thank M. Shaposhnikov for helpful discussions and R. Abramo and R. Opher 
for useful comments on this work. This work was supported by
Universidad de Buenos Aires, Fundaci\'{o}n Antorchas, Conicet and Agencia
Nacional de Promoci\'{o}n Cient\'{i}fica y T\'{e}cnica, under contract
ANPCyT 03/05229.

\section{Appendix 1: Self consistent scalar electrodynamics: closed time
path two particle irreducible effective action and {\em 1/N }approximation}

In this appendix we derive the self-consistent, causal equations for the
scalar electrodynamics, in the large {\em N} approximation, using the closed
time path formulation of quantum field theory in a Friedmann Robertson
Walker universe \cite{esteban1,steve}. This formulation helps to correctly
identify the Feynmann graphs that contribute to each order in the $1/N$
expansion of the equations.

The starting point for its derivation is the action for a set of {\em N}
charged scalar field coupled to the electromagnetic field, which in curved
spacetime reads 
\begin{eqnarray}
S_F\left( \phi _i,\phi _i^{\dagger },A_\mu \right) &=&-\int d^4x\sqrt{-g}
\left\{ g^{\mu \nu }\left( \partial _\mu -ieA_\mu \right) \phi _i\left(
\partial _\nu +ieA_\nu \right) \phi _i^{\dagger }+\left( m^2+\chi R^2\right)
\phi _i\phi _i^{\dagger }\right.  \label{aaa} \\
&&+\left. \frac 14g^{\mu \alpha }g^{\nu \beta }F_{\alpha \beta }F_{\mu \nu }-
\frac 1{2\zeta }g^{\mu \alpha }g^{\nu \beta }\partial _\mu A_\alpha \partial
_\nu A_\beta \right\}  \nonumber
\end{eqnarray}
where $x=\left( t,\bar r\right) $, $e$ is the electric charge, $g^{\mu \nu }$
the metric tensor, $\sqrt{-g}$ the determinant of the spacetime metric, $
\chi $ the coupling constant of the scalar field to gravity and $\zeta $ a
gauge fixing constant. Sum over repeated greek indices is assumed. The
complex field can be expressed in terms of real ones as $\phi =\left[ \phi
_1+i\phi _2\right] /\sqrt{2}$ and $\phi ^{\dagger }=\left[ \phi _1-i\phi _2
\right] /\sqrt{2}$. To implement the large {\em N} limit it is convenient to
re-scale the fields and coupling constant as 
\begin{equation}
\phi _n^i\rightarrow \sqrt{N}\phi _n^i,\quad A_\mu \rightarrow \sqrt{N}A_\mu
,\quad e\rightarrow \frac e{\sqrt{N}}  \label{aab}
\end{equation}
obtaining 
\begin{eqnarray}
S_F^N &=&-N\int d^4x\sqrt{-g}\left\{ \frac 12g^{\mu \nu }\partial _\mu \phi
_1^i\partial _\nu \phi _2^i+\frac 12\left( m^2+\chi R^2\right) \left( \phi
_1^{i2}+\phi _2^{n2}\right) \right.  \label{aac} \\
&&+g^{\mu \nu }eA_\mu \left( \partial _\nu \phi _1^n\phi _2^n-\phi
_1^n\partial _\nu \phi _2^n\right) +\frac{e^2}2g^{\mu \nu }A_\mu A_\nu
\left( \phi _1^{n2}+\phi _2^{n2}\right)  \nonumber \\
&&+\left. \frac 14g^{\mu \alpha }g^{\nu \beta }F_{\alpha \beta }F_{\mu \nu }-
\frac 1{2\zeta }g^{\mu \alpha }g^{\nu \beta }\partial _\mu A_\alpha \partial
_\nu A_\beta \right\}  \nonumber
\end{eqnarray}
which means that each vertex appears $N$ times in the interaction.

\subsection{Closed time path generating functional for N scalar fields in
curved spacetime}

The closed time path generating function for the n-point functions is
obtained by coupling the scalar and electromagnetic fields to sources $%
J_{a}\left( x\right) $ in each of the time paths 
\begin{eqnarray}
Z\left[ J,K\right] &=&{\cal Z}_{G}\int \prod_{n=1}^{N}DA_{\pm }^{\mu }D\phi
_{1\pm }^{n}D\phi _{2\pm }^{n}\exp \left\{ iS_{F}^{N}\left[ \phi _{1\pm
}^{n},\phi _{2\pm }^{n},A_{\pm }^{\mu }\right] \right.  \nonumber \\
&&+i\int d^{4}x\sqrt{-g}C^{ab}\left[ J_{a}^{in}\left( x\right) \phi
_{ib}^{n}\left( x\right) +J_{a}^{\mu }\left( x\right) A_{\mu b}\left(
x\right) \right]  \nonumber \\
&&+\frac{i}{2}\int d^{4}x\sqrt{-g}\int d^{4}x^{\prime }\sqrt{-g^{\prime }}%
C^{ab}C^{de}  \nonumber \\
&&\times \left[ K_{ad}^{ijnm}\left( x,x^{\prime }\right) \phi
_{b}^{in}\left( x\right) \phi _{e}^{jm}\left( x\right) +K_{ad}^{\mu \nu
}\left( x,x^{\prime }\right) A_{\mu b}\left( x\right) A_{\nu e}\left(
x^{\prime }\right) \right]  \label{aad} \\
&&+\frac{i}{2}\int d^{4}x\sqrt{-g}\int d^{4}x^{\prime }\sqrt{-g^{\prime }}%
C^{ab}C^{de}  \nonumber \\
&&\times \left. K_{ad}^{in\mu }\left( x,x^{\prime }\right) \left[ \phi
_{b}^{in}\left( x\right) A_{\mu e}\left( x^{\prime }\right) +\phi
_{e}^{in}\left( x^{\prime }\right) A_{\mu b}\left( x^{\prime }\right) \right]
\right\}  \nonumber \\
&\equiv &\exp \left[ iW\left( J,K\right) \right]  \nonumber
\end{eqnarray}
with 
\begin{equation}
S_{F}^{N}\left[ \phi _{1\pm }^{n},\phi _{2\pm }^{n},A_{\pm }^{\mu }\right]
=S_{F}^{N}\left[ \phi _{1+}^{n},\phi _{2+}^{n},A_{+}^{\mu }\right] -S_{F}^{N}
\left[ \phi _{1-}^{n},\phi _{2-}^{n},A_{-}^{\mu }\right]  \label{aae}
\end{equation}
where the plus (minus) sign corresponds to forward (backward) time
direction, $n,m=1\ldots N$. and $i,j=1$ or $2$ and ${\cal Z}_{G}$ is an
overall constant factor which arose from the integration of the ghost
fields included in the action to ensure that the result is gauge invariant.
As we are dealing with Abelian gauge fields, that integral decouples from
the others. From now on we omit ${\cal Z}_{G}$ in the forthcoming
calculations. $C^{ab}$ is the closed time path metric which reads 
\begin{eqnarray}
C_{11}&=&1  \nonumber \\
C_{22}&=&-1  \label{aaf} \\
C_{12}=C_{21}&=&0
\end{eqnarray}

The generating function for {\em connected} n-point functions is given by

\begin{equation}
W=-i\ln Z\left[ J,K\right]  \label{aag}
\end{equation}
From it we define the mean values 
\begin{equation}
\hat{\phi}_{ia}^{n}\left( x\right) =C^{ab}\frac{\delta W}{\delta
J_{ib}^{n}\left( x\right) },\quad \hat{A}_{\mu a}\left( x\right) =C^{ab}%
\frac{\delta W}{\delta J_{b}^{\mu }\left( x\right) }  \label{aah}
\end{equation}
and two point functions 
\begin{equation}
\hat{\phi}_{ia}^{n}\left( x\right) \hat{\phi}_{jb}^{m}\left( x^{\prime
}\right) +G_{ijab}^{nm}\left( x,x^{\prime }\right) =2C^{ar}C^{bs}\frac{%
\delta W}{\delta K_{rs}^{ijnm}\left( x,x^{\prime }\right) }  \label{aai}
\end{equation}
\begin{equation}
\hat{A}_{\mu a}\left( x\right) \hat{A}_{\nu b}\left( x^{\prime }\right)
+D_{\mu \nu ab}\left( x,x^{\prime }\right) =2C^{ar}C^{bs}\frac{\delta W}{%
\delta K_{rs}^{\mu \nu }\left( x,x^{\prime }\right) }  \label{aaj}
\end{equation}
\begin{equation}
\frac{1}{2}\left[ \hat{\phi}_{ia}^{n}\left( x\right) \hat{A}_{\mu b}\left(
x^{\prime }\right) +\hat{\phi}_{ia}^{n}\left( x^{\prime }\right) \hat{A}%
_{\mu b}\left( x\right) \right] +H_{iab}^{n\mu }\left( x,x^{\prime }\right)
=2C^{ar}C^{bs}\frac{\delta W}{\delta K_{irs}^{n\mu }\left( x,x^{\prime
}\right) }  \label{aak}
\end{equation}

\subsection{2PI\ Effective Action}

The closed time path 2PI effective action is defined as 
\begin{eqnarray}
\Gamma &=&W-C^{ab}\int d^{4}x\sqrt{-g}\hat{\phi}_{a}^{in}\left( x\right)
J_{b}^{in}\left( x\right) -C^{ab}\int d^{4}x\sqrt{-g}\hat{A}_{\mu a}\left(
x\right) J_{b}^{\mu }\left( x\right)  \nonumber \\
&&-\frac{1}{2}C^{ar}C^{bs}\int d^{4}x\sqrt{-g}\int d^{4}x^{\prime }\sqrt{%
-g^{\prime }}  \nonumber \\
&&\times K_{ijab}^{nm}\left( x,x^{\prime }\right) \left[ \hat{\phi}%
_{ir}^{n}\left( x\right) \hat{\phi}_{js}^{m}\left( x^{\prime }\right)
+G_{ijrs}^{nm}\left( x,x^{\prime }\right) \right]  \label{aal} \\
&&-\frac{1}{2}C^{ar}C^{bs}\int d^{4}x\sqrt{-g}\int d^{4}x^{\prime }\sqrt{%
-g^{\prime }}  \nonumber \\
&&\times K_{ab}^{\mu \nu }\left( x,x^{\prime }\right) \left[ \hat{A}_{\mu
r}\left( x\right) \hat{A}_{\nu s}\left( x^{\prime }\right) +D_{\mu \nu
rs}\left( x,x^{\prime }\right) \right]  \nonumber
\end{eqnarray}
where we have neglected the terms with mixed mean values, because as we
shall work with a system for which the mean values of the fields vanish,
these terms will vanish too for they couple through the mean values.

From this action we define the sources 
\begin{eqnarray}
\frac{\delta \Gamma }{\delta \hat{\phi}_{ia}^{n}\left( x\right) } &=&-C^{ab}%
\sqrt{-g}J_{b}^{in}\left( x\right)  \label{aam} \\
&&-\frac{1}{2}C^{ar}C^{bs}\int d^{4}x^{\prime }\sqrt{-g}\sqrt{-g^{\prime }}%
\left[ K_{ijbr}^{nm}\left( x,x^{\prime }\right) +K_{ijrb}^{nm}\left(
x,x^{\prime }\right) \right] \hat{\phi}_{js}^{m}\left( x^{\prime }\right) 
\nonumber \\
&&-\frac{1}{2}C^{ar}C^{bs}\int d^{4}x^{\prime }\sqrt{-g^{\prime }}\left[
K_{irb}^{\mu n}\left( x,x^{\prime }\right) +K_{ibr}^{\mu n}\left(
x,x^{\prime }\right) \right] \hat{A}_{\mu s}\left( x^{\prime }\right) 
\nonumber
\end{eqnarray}
\begin{eqnarray}
\frac{\delta \Gamma }{\delta \hat{A}_{\mu a}\left( x\right) } &=&-C^{ab}%
\sqrt{-g}J_{b}^{\mu }\left( x\right)  \label{aan} \\
&&-\frac{1}{2}C^{ar}C^{bs}\int d^{4}x^{\prime }\sqrt{-g}\sqrt{-g^{\prime }}%
\left[ K_{br}^{\mu \nu }\left( x,x^{\prime }\right) +K_{rb}^{\mu \nu }\left(
x,x^{\prime }\right) \right] \hat{A}_{\nu s}\left( x^{\prime }\right) 
\nonumber \\
&&-\frac{1}{2}C^{ar}C^{bs}\int d^{4}x^{\prime }\sqrt{-g^{\prime }}\left[
K_{irb}^{\mu n}\left( x,x^{\prime }\right) +K_{ibr}^{\mu n}\left(
x,x^{\prime }\right) \right] \hat{\phi}_{is}^{n}\left( x^{\prime }\right) 
\nonumber
\end{eqnarray}
\begin{equation}
\frac{\delta \Gamma }{\delta G_{rs}^{ijnm}\left( x,x^{\prime }\right) }=-%
\frac{1}{2}C^{ar}C^{bs}\sqrt{-g}\sqrt{-g^{\prime }}K_{ijab}^{nm}\left(
x,x^{\prime }\right)  \label{aao}
\end{equation}
\begin{equation}
\frac{\delta \Gamma }{\delta D_{\mu \nu rs}\left( x,x^{\prime }\right) }=-%
\frac{1}{2}C^{ar}C^{bs}\sqrt{-g}\sqrt{-g^{\prime }}K_{ab}^{\mu \nu }\left(
x,x^{\prime }\right)  \label{aap}
\end{equation}
Inverting these expressions and replacing them in the Effective Action and
shifting the fields by 
\begin{eqnarray}
\phi _{a}^{in}\left( x\right) &=&\hat{\phi}_{a}^{in}\left( x\right) +\varphi
_{a}^{in}\left( x\right)  \label{aaq} \\
A_{\mu r}\left( x\right) &=&\hat{A}_{\mu r}\left( x\right) +a_{\mu r}\left(
x\right)  \label{aaqa}
\end{eqnarray}
we obtain

\begin{eqnarray}
&&\Gamma \left. =\right. C^{ap}C^{sq}\int d^{4}x\int d^{4}x^{\prime }\left[ 
\frac{\delta \Gamma }{\delta G_{pq}^{ijnm}\left( x,x^{\prime }\right) }%
G_{rs}^{ijnm}\left( x,x^{\prime }\right) +\frac{\delta \Gamma }{\delta
D_{\mu \nu pq}\left( x,x^{\prime }\right) }D_{\mu \nu as}\left( x,x^{\prime
}\right) \right]  \nonumber \\
&&-i\ln \int_{CTP}D\varphi Da_{\mu }\exp \left\{ iS_{F}^{N}-iC^{ap}\int
d^{4}x\left[ \frac{\delta \Gamma }{\delta \hat{\phi}_{p}^{in}\left( x\right) 
}\varphi _{a}^{in}\left( x\right) +\frac{\delta \Gamma }{\delta \hat{A}_{\mu
p}\left( x\right) }\varphi _{a}^{in}\left( x\right) \right] \right.
\label{aar} \\
&&-\left. iC^{ap}C^{sq}\int d^{4}x\int d^{4}x^{\prime }\left[ \frac{\delta
\Gamma }{\delta G_{pq}^{ijnm}\left( x,x^{\prime }\right) }\varphi
_{s}^{jm}\left( x^{\prime }\right) \varphi _{a}^{in}\left( x\right) +\frac{%
\delta \Gamma }{\delta D_{\mu \nu pq}\left( x,x^{\prime }\right) }\varphi
_{a}^{\mu }\left( x\right) a_{s}^{\nu }\left( x^{\prime }\right) \right]
\right\}  \nonumber
\end{eqnarray}

This expression can be interpreted as an implicit equation whose solution is
the Effective Action. Formally this solution can be written as \cite{jackiw} 
\begin{equation}
\Gamma =S_{F}^{N}-i\ln \det \left[ G_{1}^{-1}\right] -i\ln \det \left[
G_{2}^{-1}\right] -i\ln \det \left[ D_{\mu \nu }^{-1}\right] +\Gamma _{2}
\label{aas}
\end{equation}
where $\Gamma _{2}$ is $-i\hbar $ times all the two particle irreducible
vacuum graphs with lines given by $G_{iab}$ and vertices given by a shifted
action $S_{int}$ given by 
\begin{eqnarray}
S_{int}\left[ \varphi \right] &=&S\left[ \hat{\phi}+\varphi \right] -S\left[ 
\hat{\phi}\right] -C^{ab}\int d^{4}x\frac{\delta S}{\delta \phi _{a}}\left[ 
\hat{\phi}\right] \varphi _{b}  \label{aat} \\
&&-C^{ab}C^{a^{\prime }b^{\prime }}\int d^{4}x\int d^{4}x^{\prime }\frac{%
\delta ^{2}S}{\delta \phi _{a}\left( x\right) \delta \phi _{a^{\prime
}}\left( x^{\prime }\right) }\left[ \hat{\phi}\right] \varphi _{b}\left(
x\right) \varphi _{b^{\prime }}\left( x^{\prime }\right)  \nonumber
\end{eqnarray}
Writing 
\begin{eqnarray}
{\cal G}_{nab}^{ij}\left( x,x^{\prime }\right) &=&\frac{\delta ^{2}S_{F}}{%
\delta \phi _{na}^{i}\left( x\right) \delta \phi _{nb}^{j}\left( x^{\prime
}\right) }  \label{aau} \\
{\cal D}_{ab}^{\mu \nu }\left( x,x^{\prime }\right) &=&\frac{\delta ^{2}S_{F}
}{\delta A_{\mu a}\left( x\right) \delta A_{\nu b}\left( x^{\prime }\right) }
,  \label{aav}
\end{eqnarray}
performing the scalings 
\begin{eqnarray}
G_{1ab}^{ij} &\rightarrow &\frac{1}{N}G_{1ab}\delta ^{ij}  \nonumber \\
G_{2ab}^{ij} &\rightarrow &\frac{1}{N}G_{2ab}\delta ^{ij}  \nonumber \\
D_{\mu \nu }^{ab}\left( x,x^{\prime }\right) &\rightarrow &\frac{1}{N}D_{\mu
\nu }^{ab}\left( x,x^{\prime }\right)  \label{aaw} \\
{\cal G}_{1ab}^{ij}\left( x,x^{\prime }\right) &\rightarrow &N{\cal G}
_{1ab}\left( x,x^{\prime }\right) \delta ^{ij}  \nonumber \\
{\cal G}_{2ab}^{ij}\left( x,x^{\prime }\right) &\rightarrow &N{\cal G}
_{2ab}\left( x,x^{\prime }\right) \delta ^{ij}  \nonumber \\
{\cal D}_{ab}^{\mu \nu }\left( x,x^{\prime }\right) &\rightarrow &N{\cal D}
_{ab}^{\mu \nu }\left( x,x^{\prime }\right)  \nonumber
\end{eqnarray}
and taking into account that each loop of scalar field counts $N$, that each
trace over scalar field indices counts $N$ and that there is an overall
factor of $N$ in the action, the effective action reads

\begin{eqnarray}
\Gamma &=&\frac{i}{2}N\ln \det \left[ \left( G_{1}\right) ^{-1}\right] +
\frac{i}{2}N\ln \det \left[ \left( G_{2}\right) ^{-1}\right] +\frac{i}{2}\ln
\det \left[ \left( D^{\mu \nu }\right) ^{-1}\right]  \nonumber \\
&+&\frac{N^{2}}{2}\int d^{4}x\sqrt{-g}\int d^{4}x^{\prime }\sqrt{-g^{\prime }
}{\cal G}_{1ab}\left( x,x^{\prime }\right) G_{1ab}\left( x,x^{\prime }\right)
\nonumber \\
&+&\frac{N^{2}}{2}\int d^{4}x\sqrt{-g}\int d^{4}x^{\prime }\sqrt{-g^{\prime }
}{\cal G}_{2ab}\left( x,x^{\prime }\right) G_{2ab}\left( x,x^{\prime }\right)
\nonumber \\
&&+\frac{N}{2}\int d^{4}x\sqrt{-g}\int d^{4}x^{\prime }\sqrt{-g^{\prime }}
{\cal D}_{ab}^{\mu \nu }\left( x,x^{\prime }\right) D_{\mu \nu }^{ab}\left(
x,x^{\prime }\right)  \label{aax} \\
&-N&\frac{e^{2}}{2}C^{abcd}\int d^{4}x\sqrt{-g}g^{\mu \nu }(x)D_{\mu \nu
}^{ab}\left( x,x\right) \left[ G_{1cd}\left( x,x\right) +G_{2cd}\left(
x,x\right) \right]  \nonumber \\
&+&iN\frac{e^{2}}{2}C^{abc}C^{a^{\prime }b^{\prime }c^{\prime }}\int d^{4}x%
\sqrt{-g}\int d^{4}x^{\prime }\sqrt{-g^{\prime }}D_{\mu \nu }^{aa^{\prime
}}\left( x,x^{\prime }\right) g^{\mu \alpha }\left( x\right) g^{\nu \beta
}\left( x^{\prime }\right)  \nonumber \\
&&\times \left[ G_{1bb^{\prime }}\left( x,x^{\prime }\right) 
\stackrel{\leftrightarrow}{\partial }_{\alpha }
\stackrel{\leftrightarrow}{\partial }_{\beta }^{\prime }G_{2cc^{\prime
}}\left( x,x^{\prime }\right) \right]  \nonumber
\end{eqnarray}
with $G^{ab}$ the propagators for the scalar field, and $D_{ab}^{\mu \nu }$
the propagators for the electromagnetic field. 

It is easy to check that no other loops contribute to the 2PI effective
action to this order in $1/N$. Indeed consider a graph that is order 
$N^{0}$, adding a new scalar line that maintains the 2PI condition 
amounts to add
two vertices (and hence to multiply by $N^{-2}$), one propagator (another 
$N^{-1}$) and one momentum integration (a factor of $N$), 
which gives a graph
that is order $N^{-2}$. Adding a new photon line also amounts to add two
vertices, one propagator and one loop integration, so the new graph is 
again order $N^{-2}$. As described in Section III, $G_{21}\left(
x,x^{\prime }\right) $ is the positive frequency propagator and $%
G_{12}\left( x,x^{\prime }\right) $ the negative frequency one. Their
spatial Fourier transforms read 
\begin{eqnarray}
G_{21}\left( x,x^{\prime }\right) &=&\int \frac{d^{3}\kappa }{\left( 2\pi
\right) ^{3/2}}u_{\kappa }\left( t\right) u_{\kappa }^{\ast }\left(
t^{\prime }\right) \exp \left[ i\bar{\kappa}.\left( \bar{r}-\bar{r}^{\prime
}\right) \right]  \label{aaxx} \\
G_{12}\left( x,x^{\prime }\right) &=&\int \frac{d^{3}\kappa }{\left( 2\pi
\right) ^{3/2}}u_{\kappa }^{\ast }\left( t\right) u_{\kappa }\left(
t^{\prime }\right) \exp \left[ i\bar{\kappa}.\left( \bar{r}-\bar{r}^{\prime
}\right) \right]  \label{aaxy}
\end{eqnarray}
where $u_{\kappa }\left( t\right) $ are positive frequency modes, that are
solutions to the Klein Gordon equation for the scalar field.

In the next subsections we shall find the evolution equations for the
different two point functions involved in our calculations.

\subsection{Equation for the scalar field propagator}

The equation for the scalar field propagator is obtained by taking the
functional derivative of the effective action with respect to the desired
two point function, i.e. 
\begin{eqnarray}
\frac{\delta \Gamma }{\delta G_{1ab}} &=&-\frac{i}{2}NG_{ab}^{-1}+\frac{1}{2}
N{\cal G}_{1ab}\left( x,x^{\prime }\right) -i\frac{e^{2}}{2}g^{\mu \nu
}\left( x\right) C^{abcd}D_{\mu \nu }^{ab}\left( x,x^{\prime }\right) \delta
\left( x-x^{\prime }\right)  \label{aay} \\
&&+\frac{e^{2}}{2}C^{acd}C^{bc^{\prime }d^{\prime }}D_{\mu \nu }^{cc^{\prime
}}\left( x,x^{\prime }\right) \overline{\partial }_{\alpha }\overline{
\partial }_{\beta }^{\prime }G_{2dd^{\prime }}\left( x,x^{\prime }\right)
\left. =\right. 0  \nonumber
\end{eqnarray}
from where, to leading order in $1/N$ we have 
\begin{equation}
iG_{ab}^{-1}\left( x,x^{\prime }\right) ={\cal G}_{1ab}\left( x,x^{\prime
}\right)  \label{aaz}
\end{equation}
Replacing the expression for ${\cal G}_{1ab}\left( x,x^{\prime }\right) $ we
obtain

\begin{equation}
\left\{ \frac{1}{\sqrt{-g}}\partial _{\mu }\left( \sqrt{-g}g^{\mu \nu
}\right) \partial _{\nu }+g^{\mu \nu }\partial _{\mu }\partial _{\nu
}-\left( m^{2}+\chi {\cal R}\right) \right\} C_{ab}G^{bc}\left( x,x^{\prime
}\right) =\frac{i}{\sqrt{-g}}\delta _{a}^{c}\delta \left( x-x^{\prime
}\right)  \label{aba}
\end{equation}
where the delta function is a coordinate function. We see that to this order
in $1/N$, the equations for the scalar and electromagnetic potential
decouple and therefore there is no back-reaction of the electromagnetic field
over its sources.

\subsection{Equation for the electromagnetic field propagator}

It is given by 
\begin{eqnarray}
\frac{\delta \Gamma }{\delta D_{\mu \nu }} &=&-\frac i2ND_{ab}^{\mu \nu -1}+
\frac 12N{\cal D}_{ab}^{\mu \nu }\left( x,x^{\prime }\right) -N\frac{e^2}2
\left[ G_{1ab}\left( x,x\right) +G_{2ab}\left( x,x\right) \right]  \nonumber
\\
&&+iN\frac{e^2}2C^{acd}C^{bc^{\prime }d^{\prime }}g^{\mu \alpha }\left(
x\right) g^{\nu \beta }\left( x^{\prime }\right) \left[ G_{1cc^{\prime
}}\left( x,x^{\prime }\right) \overline{\partial }_\alpha \overline{\partial 
}_\beta ^{\prime }G_{2dd^{\prime }}\left( x,x^{\prime }\right) \right]
\label{abb} \\
&=&0  \nonumber
\end{eqnarray}
where the factor $N$ in the term $ND_{ab}^{\mu \nu -1}$ comes from the
scaling of the inverse of the propagator. This is already the equation to
leading order in the $1/N$ expansion. Replacing ${\cal D}_{ab}^{\mu \nu
}\left( x,x^{\prime }\right) $: 
\begin{eqnarray}
{\cal D}_{ab}^{\mu \nu }\left( x,x^{\prime }\right) &=&\sqrt{-g\left(
x\right) }\left\{ g^{\mu \nu }\left( x\right) \Box _x+\left( 1-\frac 1\zeta
\right) g^{\mu \alpha }\left( x\right) g^{\nu \beta }\left( x\right)
\partial _\alpha \partial _\beta \right.  \nonumber \\
&&-\left. \frac 1{\sqrt{-g}}\partial _\alpha \left[ \sqrt{-g}g^{\mu \alpha
}g^{\nu \beta }\right] A_\beta F_{\mu \nu }-\frac 1{\xi \sqrt{-g}}\partial
_\mu \left[ \sqrt{-g}g^{\mu \alpha }g^{\nu \beta }\right] A_\alpha \partial
_\nu A_\beta \right\} C_{ab}  \label{abc} \\
&\equiv &\Delta _{ab}^{\mu \nu }\left( x,x^{\prime }\right)  \nonumber
\end{eqnarray}
The equation for the propagators for the electromagnetic field then reads 
\begin{eqnarray}
&&\Delta _{ab}^{\mu \nu }\left( x,x^{\prime }\right) D_{\nu \rho
}^{bf}\left( x,x^{\prime }\right) -e^2g^{\mu \nu }\left( x\right) \left[
G_{1ab}\left( x,x\right) +G_{2ab}\left( x,x\right) \right] D_{\nu \rho
}^{bf}\left( x,x^{\prime }\right)  \nonumber \\
&&+ie^2C^{acd}C^{bc^{\prime }d^{\prime }}\int dx^{\prime \prime }\sqrt{%
-g\left( x^{\prime \prime }\right) }g^{\mu \alpha }\left( x\right) g^{\nu
\beta }\left( x^{\prime \prime }\right)  \label{abd} \\
&&\times \left[ G_{1cc^{\prime }}\left( x,x^{\prime \prime }\right) 
\overline{\partial }_\alpha \overline{\partial }_\beta ^{\prime
}G_{2dd^{\prime }}\left( x,x^{\prime \prime }\right) \right] D_{\nu \rho
}^{bf}\left( x^{\prime \prime },x^{\prime }\right)  \nonumber \\
&&\left. =\right. i\delta _{\mu \rho }\delta _{af}\frac 1{\sqrt{-g\left(
x\right) }}\delta \left( x-x^{\prime }\right)  \nonumber
\end{eqnarray}

\subsection{Changing to conformal time}

By changing to conformal time $d\eta =dt/a\left( t\right) $, using
$\tau = H\eta$ and performing
the scalings 
\begin{equation}
G_{iab}\left( x,x^{\prime }\right) \rightarrow \frac{G_{iab}\left(
x,x^{\prime }\right) }{a\left( \tau \right) a\left( \tau ^{\prime }\right) }
\label{abe}
\end{equation}
it can be checked by a straightforward calculation that the equation for the
propagators read 
\begin{equation}
\left\{ \Box _x - \frac{m^2}{H^2}a^2\left( \tau \right) -
\left( 6\chi -1\right) \frac{\ddot a\left( \tau \right) }{a\left( 
\tau \right) }\right\}
C_{ab}G^{bc}\left( x,x^{\prime }\right) =i\delta _a^c\delta \left(
x-x^{\prime }\right)  \label{abf}
\end{equation}
and 
\begin{eqnarray}
&&\left[ \tau ^{\mu \nu }\Box _x+\left( 1-\frac 1\zeta \right) \tau ^{\mu
\alpha }\tau ^{\nu \beta }\partial _\alpha \partial _\beta \right]
C_{ab}D_{\nu \rho }^{bf}\left( x,x^{\prime }\right)  \nonumber \\
&&-e^2\Gamma _{cd}^{\mu \nu }\left( x,x^{\prime }\right) D_{\nu \rho
}^{bf}\left( x,x^{\prime }\right) +ie^2C^{acd}C^{bc^{\prime }d^{\prime
}}\int dx^{\prime \prime }\Sigma _{cc^{\prime },dd^{\prime }}^{\mu \nu}
\left( x,x^{\prime\prime} \right)
D_{\nu \rho }^{bf}\left( x^{\prime \prime },x^{\prime }\right)  \label{abg}
\\
&=&i\delta _{\mu \rho }\delta _{af}\delta \left( x-x^{\prime }\right) 
\nonumber
\end{eqnarray}
where 
\begin{equation}
\Box _x=\eta ^{\alpha \beta }\partial _\alpha \partial _\beta =-\partial
_0^2+\nabla ^2  \label{abh}
\end{equation}
and

\begin{equation}
\Gamma _{cd}^{\mu \nu }\left( x,x\right) \equiv \eta ^{\mu \nu } 
\left[ G_{1cd}\left( x,x\right) +G_{2cd}\left( x,x\right) \right]
\label{abi}
\end{equation}

\begin{equation}
\Sigma _{cc^{\prime },dd^{\prime }}^{\mu \nu }\equiv \eta ^{\mu \alpha
}(x)\eta ^{\nu \beta }(x")\left[ G_{1cc^{\prime }}\left( x,x"\right) 
\overline{\partial }_{\alpha }\overline{\partial }_{\beta }^{\prime \prime
}G_{2dd^{\prime }}\left( x,x"\right) \right]  \label{abj}
\end{equation}
To avoid over-notation, we have also used $x$ as the dimensionless variables.

\subsection{Equations for the retarded and Hadamard two point functions of
the electromagnetic four potential}

In this section we shall find the equations for the retarded and Hadamard
propagators. The latter encodes all the information about the state of the
electromagnetic field while the former evolves the initial conditions. For
the electromagnetic field we have

\begin{eqnarray}
&&\left[ \eta ^{\mu \nu }\Box _x+\left( 1-\frac 1\zeta \right) \partial ^\mu
\partial ^\nu -e^2\Gamma _{11}^{\mu \nu }\left( x,x\right) \right] D_{\nu
\gamma }^{11}\left( x,x^{\prime }\right)  \label{abk} \\
&&+ie^2\int dx"\Sigma _{11,11}^{\mu \nu }\left( x,x"\right) D_{\nu \gamma
}^{11}\left( x",x^{\prime }\right)  \nonumber \\
&&-ie^2\int dx"\Sigma _{12,12}^{\mu \nu }\left( x,x"\right) D_{\nu \gamma
}^{21}\left( x",x^{\prime }\right) \left. =\right. i\delta _\gamma ^\mu
\delta \left( x-x^{\prime }\right)  \nonumber
\end{eqnarray}

\begin{eqnarray}
&&\left[ \eta ^{\mu \nu }\Box _x+\left( 1-\frac 1\zeta \right) \partial ^\mu
\partial ^\nu -e^2\Gamma _{11}^{\mu \nu }\left( x,x\right) \right] D_{\nu
\gamma }^{12}\left( x,x^{\prime }\right)  \label{abl} \\
&&+ie^2\int dx"\Sigma _{11,11}^{\mu \nu }\left( x,x"\right) D_{\nu \gamma
}^{12}\left( x",x^{\prime }\right)  \nonumber \\
&&-ie^2\int dx"\Sigma _{12,12}^{\mu \nu }\left( x,x"\right) D_{\nu \gamma
}^{22}\left( x",x^{\prime }\right) \left. =\right. 0  \nonumber
\end{eqnarray}

\begin{eqnarray}
&&\left[ \eta ^{\mu \nu }\Box _x+\left( 1-\frac 1\zeta \right) \partial ^\mu
\partial ^\nu -e^2\Gamma _{22}^{\mu \nu }\left( x,x\right) \right] D_{\nu
\gamma }^{21}\left( x,x^{\prime }\right)  \label{abm} \\
&&+ie^2\int dx"\Sigma _{21,21}^{\mu \nu }\left( x,x"\right) D_{\nu \gamma
}^{11}\left( x",x^{\prime }\right)  \nonumber \\
&&-ie^2\int dx"\Sigma _{22,22}^{\mu \nu }\left( x,x"\right) D_{\nu \gamma
}^{21}\left( x",x^{\prime }\right) \left. =\right. 0  \nonumber
\end{eqnarray}

\begin{eqnarray}
&&\left[ \eta ^{\mu \nu }\Box _x+\left( 1-\frac 1\zeta \right) \partial ^\mu
\partial ^\nu -e^2\Gamma _{22}^{\mu \nu }\left( x,x\right) \right] D_{\nu
\gamma }^{22}\left( x,x^{\prime }\right)  \label{abn} \\
&&+ie^2\int dx"\Sigma _{21,21}^{\mu \nu }\left( x,x"\right) D_{\nu \gamma
}^{12}\left( x",x^{\prime }\right)  \nonumber \\
&&-ie^2\int dx"\Sigma _{22,22}^{\mu \nu }\left( x,x"\right) D_{\nu \gamma
}^{22}\left( x",x^{\prime }\right) \left. =\right. -i\delta _\gamma ^\mu
\delta \left( x-x^{\prime }\right)  \nonumber
\end{eqnarray}

By substracting the first two (or the last two) equation and defining $%
\Sigma _{ret}^{\mu \nu }\equiv \Sigma _{11,11}^{\mu \nu }-\Sigma
_{12,12}^{\mu \nu }\equiv \Sigma _{21,21}^{\mu \nu }-\Sigma _{22,22}^{\mu
\nu }$, we get

\begin{eqnarray}
&&\left[ \eta ^{\mu \nu }\Box _x+\left( 1-\frac 1\zeta \right) \partial ^\mu
\partial ^\nu -e^2\Gamma _{11}^{\mu \nu }\left( x,x\right) \right] D_{\nu
\gamma }^{ret}\left( x,x^{\prime }\right)  \label{aboz} \\
&&+ie^2\int dx"\Sigma _{ret}^{\mu \nu }\left( x,x"\right) D_{\nu \gamma
}^{ret}\left( x",x^{\prime }\right) \left. =\right. -\delta _\gamma ^\mu
\delta \left( x-x^{\prime }\right)  \nonumber
\end{eqnarray}
and by adding the homogeneous equations

\begin{eqnarray}
&&\left[ \eta ^{\mu \nu }\Box _x+\left( 1-\frac 1\zeta \right) \partial ^\mu
\partial ^\nu -e^2\Gamma _{11}^{\mu \nu }\left( x,x\right) \right] D_{1\nu
\gamma }\left( x,x^{\prime }\right)  \label{abp} \\
&&+ie^2\int dx"\Sigma _{ret}^{\mu \nu }\left( x,x"\right) D_{1\nu \gamma
}\left( x",x^{\prime }\right) \left. =\right. -\frac{e^2}2\int dx"\Sigma
_1^{\mu \nu }\left( x,x"\right) D_{\nu \gamma }^{adv}\left( x",x^{\prime
}\right)  \nonumber
\end{eqnarray}
where $\Sigma _1^{\mu \nu }\equiv \Sigma _{12,12}^{\mu \nu }+\Sigma
_{21,21}^{\mu \nu }=\Sigma _{11,11}^{\mu \nu }+\Sigma _{22,22}^{\mu \nu }$.

\subsection{Fourier transformed equations}

In this subsection we find the expressions for the spatially Fourier
transformed quantities that enter the Schwinger Dyson equations, and for the
equations themselves, in terms of the dimensionless variables defined in the
body of the paper. The starting point is the Fourier transformation of the
two point functions, i.e. 
\begin{equation}
G_{cd}\left( x,x^{\prime }\right) =\int \frac{d^3p}{\left( 2\pi \right)
^{3/2}}G_{cd}\left( p,\tau ,\tau ^{\prime }\right) \exp \left[ -i\bar p%
.\left( \bar y-\bar y^{\prime }\right) \right]  \label{laa}
\end{equation}
where in terms of the modes the functions $G_{cd}\left( p,\tau ,\tau
^{\prime }\right) $ read 
\begin{eqnarray}
G_{21}\left( p,\tau ,\tau ^{\prime }\right) &=&f_p\left( \tau \right)
f_p^{*}\left( \tau ^{\prime }\right)  \label{lab} \\
G_{12}\left( p,\tau ,\tau ^{\prime }\right) &=&f_p\left( \tau ^{\prime
}\right) f_p^{*}\left( \tau \right)  \label{lac}
\end{eqnarray}
For the non local kernel we have 
\begin{equation}
\Sigma _{ret}^{\mu \nu }\left( x,x"\right) =\int \frac{d^3p}{\left( 2\pi
\right) ^{3/2}}\int \frac{d^3q}{\left( 2\pi \right) ^{3/2}}\Sigma
_{ret}^{\mu \nu }\left( p,q,\tau ,\tau "\right) \exp \left[ i\left( \bar p+%
\bar q\right) .\left( \bar y-\bar y"\right) \right]  \label{lad}
\end{equation}
with 
\begin{equation}
\Sigma _{ret}^{00}\left( p,q,\tau ,\tau "\right) =\eta ^{00}\eta
^{00}\left\{ G_{11}^1\left( p,\tau ,\tau "\right) \bar \partial _0\bar %
\partial "_0G_{11}^2\left( q,\tau ,\tau "\right) -G_{12}^1\left( p,\tau
,\tau "\right) \bar \partial _0\bar \partial "_0G_{12}^2\left( p,\tau ,\tau
"\right) \right\}  \label{lae}
\end{equation}
\begin{equation}
\Sigma _{ret}^{0i}\left( p,q,\tau ,\tau "\right) =-i\left( q_j-p_j\right)
\eta ^{00}\eta ^{ij}\left\{ G_{11}^1\left( p,\tau ,\tau "\right) \bar %
\partial _0G_{11}^2\left( q,\tau ,\tau "\right) -G_{12}^1\left( p,\tau ,\tau
"\right) \bar \partial _0G_{12}^2\left( q,\tau ,\tau "\right) \right\}
\label{laf}
\end{equation}
\begin{equation}
\Sigma _{ret}^{i0}\left( p,q,\tau ,\tau "\right) =i\left( q_j-p_j\right)
\eta ^{00}\eta ^{ij}\left\{ G_{11}^1\left( p,\tau ,\tau "\right) \bar %
\partial "_0G_{11}^2\left( q,\tau ,\tau "\right) -G_{12}^1\left( p,\tau
,\tau "\right) \bar \partial "_0G_{12}^2\left( q,\tau ,\tau "\right) \right\}
\label{lag}
\end{equation}
\begin{eqnarray}
\Sigma _{ret}^{ij}\left( p,q,\tau ,\tau "\right) &=&\left\{
q^iq^j+p^ip^j-p^iq^j-p^jq^i\right\}  \label{lah} \\
&&\times \left\{ G_{11}^1\left( p,\tau ,\tau "\right) G_{11}^2\left( q,\tau
,\tau "\right) -G_{12}^1\left( p,\tau ,\tau "\right) G_{12}^2\left( p,\tau
,\tau "\right) \right\}  \nonumber
\end{eqnarray}
For the electromagnetic two point functions we write 
\begin{equation}
D_{\nu \gamma }\left( x,x^{\prime }\right) =\int \frac{d^3k}{\left( 2\pi
\right) ^{3/2}}D_{\nu \gamma }\left( k,\tau ,\tau ^{\prime }\right) \exp %
\left[ i\bar k.\left( \bar y-\bar y^{\prime }\right) \right]  \label{lai}
\end{equation}
so the spatially Fourier transformed equations read 
\begin{eqnarray}
&&\left[ \eta ^{\mu \nu }\left( \partial _\tau ^2-\nabla ^2\right) -\left( 1-%
\frac 1\zeta \right) \partial ^\mu \partial ^\nu +e^2\int \frac{d^3p}{\left(
2\pi \right) ^{3/2}}\Gamma _{11}^{\mu \nu }\left( p,\tau ,\tau \right) %
\right] D_{\nu \gamma }^{ret}\left( k,\tau ,\tau ^{\prime }\right)
\label{laj} \\
&&-ie^2\int d\tau "\Sigma _{ret}^{\mu \nu }\left( p,k-p,\tau ,\tau "\right)
D_{\nu \gamma }^{ret}\left( k,\tau ",\tau ^{\prime }\right) \left. =\right.
\delta _\gamma ^\mu \delta \left( \tau -\tau ^{\prime }\right)  \nonumber
\end{eqnarray}
and by adding the homogeneous equations

\begin{eqnarray}
&&\left[ \eta ^{\mu \nu }\left( \partial _\tau ^2-\nabla ^2\right) -
\left( 1-
\frac 1\zeta \right) \partial ^\mu \partial ^\nu +e^2\int \frac{d^3p}{\left(
2\pi \right) ^{3/2}}\Gamma _{11}^{\mu \nu }\left( p,\tau ,\tau \right) 
\right] D_{1\nu \gamma }\left( k,\tau ,\tau ^{\prime }\right)  \label{lak} \\
&&-ie^2\int d\tau "\Sigma _{ret}^{\mu \nu }\left( p,k-p,\tau ,\tau "\right)
D_{1\nu \gamma }\left( k,\tau ",\tau ^{\prime }\right) \left. =\right. 
\frac{
e^2}2\int d\tau "\Sigma _1^{\mu \nu }\left( p,k-p,\tau ,\tau "\right) D_{\nu
\gamma }^{adv}\left( k,\tau ^{\prime },\tau "\right)  \nonumber
\end{eqnarray}

\section{Appendix 2: Computing the kernels}

\subsection{Ultraviolet sector}

The contribution from short wavelength modes to the local term is 
\begin{equation}
\Gamma _{11}^{il\left( P\right) }\left( \tau ,\tau \right) =\eta
^{il}\int_{\Lambda }^{\infty }\frac{d^{3}p}{\left( 2\pi \right) ^{3/2}}%
\left\{ 2\left| \beta _{p}\right| ^{2}\left| f_{p}\left( \tau \right)
\right| ^{2}+\alpha _{p}\beta _{p}^{\ast }f_{p}^{2}\left( \tau \right)
+\beta _{p}\alpha _{p}^{\ast }f_{p}^{\dagger 2}\left( \tau \right) \right\}
\label{acib}
\end{equation}
where $\Lambda $ is the minimum wavenumber for which the approximation of
short wavelengths becomes valid. Replacing eqs. (\ref{xad}) and (\ref{xaf})
and performing the momentum integrals we get. 
\begin{equation}
\Gamma _{11}^{il\left( P\right) }\left( \tau ,\tau \right) \simeq \frac{9\pi 
}{32\left( 2\pi \right) ^{3/2}\Lambda ^{2}}-\frac{3}{2\left( 2\pi \right)
^{3/2}}%
\mathop{\rm Ci}%
\left[ 2\Lambda \tau \right]  \label{aciib}
\end{equation}
For the mixed non local: 
\begin{equation}
\Sigma _{ret}^{i0\left( P\right) }\left( p,k-p,\tau ,\tau "\right) \left.
=\right. 0  \label{ackb}
\end{equation}
and for the pure spatial non local 
\begin{eqnarray}
&&\Sigma _{ret}^{il\left( P\right) }\left( p,k-p,\tau ,\tau "\right) \simeq
\Theta \left( \tau -\tau ^{\prime \prime }\right) 4p^{i}p^{l}  \label{acjb}
\\
&&\left\{ 2\left| \beta _{p}\right| ^{2}\left[ f_{p}^{2}\left( \tau \right)
f_{p}^{\ast 2}\left( \tau "\right) -f_{p}^{\ast 2}\left( \tau \right)
f_{p}^{2}\left( \tau "\right) \right] \right.  \nonumber \\
&&+2\alpha _{p}\beta _{p}^{\ast }\left[ f_{p}\left( \tau \right) f_{p}^{\ast
}\left( \tau "\right) -f_{p}^{\ast }\left( \tau \right) f_{p}\left( \tau
"\right) \right] f_{p}\left( \tau \right) f_{p}\left( \tau "\right) 
\nonumber \\
&&\left. 2\beta _{p}\alpha _{p}^{\ast }\left[ f_{p}\left( \tau \right)
f_{p}^{\ast }\left( \tau "\right) -f_{p}^{\ast }\left( \tau \right)
f_{p}\left( \tau "\right) \right] f_{p}^{\ast }\left( \tau \right)
f_{p}^{\ast }\left( \tau "\right) \right\}  \nonumber
\end{eqnarray}
Replacing the modes and Bogolyubov coefficients we are left with 
\begin{eqnarray}
&&\Sigma _{ret}^{il\left( P\right) }\left( p,k-p,\tau ,\tau "\right) \left.
\simeq \right.  \label{adb} \\
&&4\Theta \left( \tau -\tau ^{\prime \prime }\right) p^{i}p^{l}\left\{ -%
\frac{9i}{64}\frac{1}{p^{6}}\sin \left[ 2p\left( \tau -\tau "\right) \right]
+\frac{3}{8p^{4}}\cos \left[ 2p\tau \right] -\frac{3}{8p^{4}}\cos \left[
2p\tau "\right] \right\}  \nonumber
\end{eqnarray}
To evaluate the momentum integral, we must take into account that we shall
be interested in the transverse component of the electromagnetic two point
function, which is obtained by taking the curl of the corresponding
equations. If we assume that the magnetic field propagates along the $z$%
-direction, then the curl picks out the components $x-x$, $x-y$ and $y-y$ of
the equation, and of these only the $x-x$ and $y-y$ are non vanishing. For
either of them we get 
\begin{eqnarray}
&&\int_{\Lambda }^{\infty }\frac{d^{3}p}{\left( 2\pi \right) ^{3/2}}\Sigma
_{ret}^{xx\left( P\right) }\left( p,k-p,\tau ,\tau "\right) \left. =\right. 
\nonumber \\
&&\eta ^{xx}\sqrt{\frac{2}{9\pi }}\left\{ -\frac{9i}{16}\frac{\sin \left[
2\Lambda \left( \tau -\tau "\right) \right] }{\Lambda }+\frac{9i}{8}\left(
\tau -\tau "\right) 
\mathop{\rm Ci}%
\left[ 2\Lambda \left( \tau -\tau "\right) \right] \right.  \label{adc} \\
&&-\left. \frac{3}{4\tau }\cos \left[ 2\Lambda \tau \right] +\frac{3}{4\tau "%
}\cos \left[ 2\Lambda \tau "\right] \right\}  \nonumber
\end{eqnarray}

We see that for both kernels the only momentum contribution is due to the
lower limit of integration and these terms must cancel against the ones that
come from the upper limit of the integral in the intermediate momenta.

For the pure spatial Hadamard kernel we keep only the terms quadratic in the
Bogolyubov coefficients, as they will give the main contribution in this
momentum interval, i.e. we have 
\begin{eqnarray}
&&\Sigma _{1}^{il\left( P\right) }\left( p,k-p,\tau ,\tau "\right) \simeq
4p^{i}p^{l}  \label{acfb} \\
&&\left\{ 2\left| \beta _{p}\right| ^{2}\left[ 2\left| f_{p}\left( \tau
\right) \right| ^{2}\left| f_{p}\left( \tau "\right) \right| +f_{p}^{\ast
2}\left( \tau "\right) f_{p}^{2}\left( \tau \right) +f_{p}^{2}\left( \tau
"\right) f_{p}^{\ast 2}\left( \tau \right) \right] \right.  \nonumber \\
&&+2\beta _{p}\alpha _{p}^{\ast }\left[ \left| f_{p}\left( \tau \right)
\right| ^{2}f_{p}^{\ast }\left( \tau "\right) +\left| f_{p}\left( \tau
"\right) \right| ^{2}f_{p}^{\ast }\left( \tau \right) \right]  \nonumber \\
&&+\left. 2\beta _{p}^{\ast }\alpha _{p}\left[ \left| f_{p}\left( \tau
"\right) \right| ^{2}f_{p}\left( \tau \right) +\left| f_{p}\left( \tau
\right) \right| ^{2}f_{p}\left( \tau "\right) \right] \right\}  \nonumber
\end{eqnarray}
and for which the same reasoning as for the retarded kernel applies. We can
conclude that the ultraviolet sector of the created particles spectrum makes
no contribution to the noise kernels. A way to understand it is to observe
that the mode functions for inflation and reheating are given by the same
Bessel function, $H_{3/2}^{\left( 2\right) }\left( z\right) $, but with
different arguments. This fact makes very small the amount of created
particles, as can be seen from the expression for $\beta _{p}$.

\subsection{Infrared sector}

The local kernel for this portion of the momenta spectrum was computed in
the body of the paper. In this part of the appendix we shall evaluate the
retarded non local kernel and show that it is negligible with respect to the
local one. The computation of the non local Hadamard kernel $\Sigma _1^{il}$
is straightforward.

For the $i-0$ components of the retarded kernel we have: 
\begin{eqnarray}
&&\Sigma _{ret}^{i0}\left( p,k-p,\tau ,\tau "\right) \left. =\right.
-i\Theta \left( \tau -\tau ^{\prime \prime }\right) \left(
k^{i}-2p^{i}\right)  \nonumber \\
&&\left\{ \left( \beta _{p}^{\ast }\alpha _{p}-\beta _{k-p}^{\ast }\alpha
_{k-p}\right) f\left( \tau \right) f\left( \tau \right) \left[ \dot{f}^{\ast
}\left( \tau "\right) f\left( \tau "\right) -f^{\ast }\left( \tau "\right) 
\dot{f}\left( \tau "\right) \right] \right.  \nonumber \\
&&+\left( \alpha _{k-p}^{\ast }\beta _{k-p}-\alpha _{p}^{\ast }\beta
_{p}\right) f^{\ast }\left( \tau \right) f^{\ast }\left( \tau \right) \left[ 
\dot{f}\left( \tau "\right) f^{\ast }\left( \tau "\right) -f\left( \tau
"\right) \dot{f}^{\ast }\left( \tau "\right) \right]  \label{acjz} \\
&&+\left. 2\left( \left| \beta _{k-p}\right| ^{2}-\left| \beta _{p}\right|
^{2}\right) f^{\ast }\left( \tau \right) f\left( \tau \right) \left[ \dot{f}%
\left( \tau "\right) f^{\ast }\left( \tau "\right) -f\left( \tau "\right) 
\dot{f}^{\ast }\left( \tau "\right) \right] \right\}  \nonumber
\end{eqnarray}
while for the pure spatial components of the same kernel we get 
\begin{eqnarray}
&&\Sigma _{ret}^{il}\left( p,k-p,\tau ,\tau "\right) \left. =\right. \Theta
\left( \tau -\tau ^{\prime \prime }\right) \left(
4p^{i}p^{l}+k^{i}k^{l}-2p^{i}k^{l}-2k^{i}p^{l}\right)  \nonumber \\
&&\left\{ \left( \left| \beta _{p}\right| ^{2}+\left| \beta _{k-p}\right|
^{2}\right) \left[ f^{2}\left( \tau \right) f^{\ast 2}\left( \tau "\right)
-f^{\ast 2}\left( \tau \right) f^{2}\left( \tau "\right) \right] \right. 
\nonumber \\
&&+\left( \alpha _{k-p}\beta _{k-p}^{\ast }+\alpha _{p}\beta _{p}^{\ast
}\right) \left[ f\left( \tau \right) f^{\ast }\left( \tau "\right) -f^{\ast
}\left( \tau \right) f\left( \tau "\right) \right] f\left( \tau \right)
f\left( \tau "\right)  \label{ackz} \\
&&+\left( \beta _{k-p}\alpha _{k-p}^{\ast }+\beta _{p}\alpha _{p}^{\ast
}\right) \left[ f\left( \tau \right) f^{\ast }\left( \tau "\right) -f^{\ast
}\left( \tau \right) f\left( \tau "\right) \right] f^{\ast }\left( \tau
\right) f^{\ast }\left( \tau "\right)  \nonumber
\end{eqnarray}

Recalling again that we shall take the curl of the equations and considering
for example, that the magnetic field propagates along the $z$ direction, we
need only the $0-x$ and $0-y$ components of the mixed kernel and the $x-x$, $%
x-y$ and $y-y$ of the pure spatial one. But except for the $x-x$ and $y-y$
of the pure spatial noise kernel, the rest of the components, by virtue of
the momentum independence of the mode function (\ref{wbn}), are odd
functions of $p$ and hence their momentum integral vanishes. We then
find the desired decoupling of the equation for the transverse part of the
pure spatial propagator. Of course the equations for longitudinal part of
the propagators are still coupled and
carry the information about charge conservation.

Performing the momentum integral of the component $\Sigma _{ret}^{xx\left(
N\right) }$ for example, we get 
\begin{eqnarray}
&&\int_\Upsilon ^\Delta \frac{d^3p}{\left( 2\pi \right) ^{3/2}}\Sigma
_{ret}^{xx\left( P\right) }\left( p,k-p,\tau ,\tau "\right) \left. \sim
\right. \eta ^{xx}\Theta \left( \tau -\tau ^{\prime \prime }\right) \frac 83%
\Delta ^2  \label{adl} \\
&&2^{3/\gamma -4}\Gamma ^2\left( \frac{3+2\gamma }{2\gamma }\right) \left( 
\frac \pi \gamma \right) \frac{z^{1/\gamma }z"^{1/\gamma }}{z_0^{5/\gamma }}
\nonumber \\
&&\left( H_{3/2\gamma }^{\left( 2\right) }\left( z\right) H_{3/2\gamma
}^{\left( 1\right) }\left( z"\right) -H_{3/2\gamma }^{\left( 2\right)
}\left( z"\right) H_{3/2\gamma }^{\left( 1\right) }\left( z\right) \right) 
\nonumber \\
&&\times \left[ H_{3/2\gamma }^{\left( 2\right) }\left( z\right)
H_{3/2\gamma }^{\left( 1\right) }\left( z"\right) +H_{3/2\gamma }^{\left(
2\right) }\left( z"\right) H_{3/2\gamma }^{\left( 1\right) }\left( z\right)
\right.  \nonumber \\
&&+\left. H_{3/2\gamma }^{\left( 2\right) }\left( z"\right) H_{3/2\gamma
}^{\left( 2\right) }\left( z\right) +H_{3/2\gamma }^{\left( 1\right) }\left(
z\right) H_{3/2\gamma }^{\left( 1\right) }\left( z"\right) \right]  \nonumber
\end{eqnarray}

Comparing the prefactor in this equation with the corresponding one in eq. 
(\ref{acnb}) we see that by virtue of the logarithmic dependence in the
latter, the non local kernel turns out to be several orders of magnitude
smaller than the local one.

\end{document}